\documentclass[useAMS,usenatbib]{mn2e}
\usepackage[dvips]{graphicx}
\usepackage[]{txfonts}
\usepackage{subfigure}

\def\lesssim{\mathbin{\lower 3pt\hbox
{$\rlap{\raise 5pt\hbox{$\char'074$}}\mathchar"7218$}}}   
\def\gtrsim{\mathbin{\lower 3pt\hbox
{$\rlap{\raise 5pt\hbox{$\char'076$}}\mathchar"7218$}}}   

\def\ba{\begin{eqnarray}}
\def\ea{\end{eqnarray}}

\newcommand       \bea          {\begin{eqnarray}}
\newcommand       \eea          {\end{eqnarray}}

\title[]{Global models of runaway accretion in white dwarf debris disks}

\author[Metzger, Rafikov, $\&$ Bochkarev]
{Brian D.~Metzger$^{1,2}$\thanks{E-mail: bmetzger, rrr, kbochkar@astro.princeton.edu}, Roman R.~Rafikov$^{1,3}$, Konstantin V.~Bochkarev$^{1,4}$\\
$^{1}$Department of Astrophysical Sciences, Peyton Hall, Princeton
  University, Princeton, NJ 08544, USA\\
$^{2}$NASA Einstein Fellow\\
$^{3}$Sloan Fellow\\
$^{4}$Department of General and Applied Physics, Moscow 
Institute of Physics and Technology, Dolgoprudny, 141700, Russia\\}

\begin{document}
\date{Received/Accepted}
\pagerange{\pageref{firstpage}--\pageref{lastpage}} \pubyear{2012}

\maketitle
\label{firstpage}


\begin{abstract}
A growing sample of young white dwarfs (WDs) with metal-enriched atmospheres are accompanied by excess infrared emission, indicating that they are encircled by a compact dusty disk of solid debris.  Such `WD debris disks' are thought to originate from the tidal disruption of asteroids or other minor bodies.  However, the precise mechanism(s) responsible for transporting matter from the disruption radius to the WD surface remains unclear, especially in those systems with the highest inferred metal accretion rates $\dot{M}_{Z} \sim 10^{8}-10^{10}$ g s$^{-1}$, which cannot be explained by Poynting-Robertson (PR) drag alone.  Here we present global time-dependent calculations of the coupled evolution of the gaseous and solid components of WD debris disks.  Solids transported inwards (initially due to PR drag) sublimate at tens of WD radii, producing a source of gas that both accretes onto the WD surface and viscously spreads outwards in radius, where it overlaps with the solid disk.  Our calculations show that if the aerodynamic coupling between the solids and gaseous disks is sufficiently strong (and/or the gas viscosity sufficiently weak), then gas builds up near the sublimation radius faster than it can viscously spread away.  Since the rate of drag-induced solid accretion increases with gas density, this results in a runaway accretion process, as predicted by Rafikov, during which the WD accretion rate reaches values orders of magnitude higher than can be achieved by PR drag alone, consistent with the highest measured values of $\dot{M}_{Z}$.  We explore the evolution of WD debris disks across a wide range of physical conditions and describe the stages of the runaway process in detail.  We also calculate the predicted distribution of observed accretion rates $\dot{M}_{Z}$, finding reasonable agreement with the current sample.  Although the conditions necessary for runaway accretion are at best marginally satisfied given the minimal level of aerodynamic drag between circular gaseous and solid disks, the presence of other stronger forms of solid-gas coupling---such as would result if the gaseous disk is only mildly eccentric---substantially increase the likelihood of runaway accretion.
\end{abstract}



\section{Introduction}  
\label{sect:intro}


A significant fraction $\sim 20$ percent of white dwarfs (WDs) with metal-enriched atmospheres (DZ type) and cooling ages $\lesssim$ 0.5 Gyr are accompanied by excess infrared (IR) emission (\citealt{Zuckerman&Becklin87}; \citealt{Graham+90}; \citealt{Kilic+05}; \citealt{vonHippel+07}; \citealt{Kilic&Redfield07}; \citealt{Farihi+09}; \citealt{Debes+10}; \citealt{Debes+11}).  This emission is well modeled as WD radiation reprocessed by an optically-thick, geometrically-thin disk of refractory material (\citealt{Jura03}; \citealt{Jura+07a,Jura+07b}).  Modeling of the IR spectral energy distribution [SED] shows that the inner edge of the disk resides at $R_{\rm in} \sim$ several tens of WD radii $R_\star$, close to where the equilibrium temperature $T \sim 1500$ K is sufficiently high to sublimate silicate dust grains. The outer edge of the disk is also well constrained by the SED at longer wavelengths, with its radius $R_{\rm out} \lesssim R_{\odot} \approx 100R_{\star}$ typically found to exceed $R_{\rm in}$ by a factor of several (e.g.~\citealt{Jura+07b}; \citealt{Jura+09a}; \citealt{Debes+11}). 

No evidence is found for disk flaring from the SEDs of most systems (cf.~\citealt{Jura+07a}), consistent with the disk material being concentrated in a thin, optically-thick layer similar to the rings of Saturn \citep{Cuzzi+10}.  High-resolution spectroscopy with IRS on the {\it Spitzer Space Telescope} reveals an emission feature at $10~\mu$m, indicating the presence of small, micron-sized silicate dust particles \citep{Jura+09b}.  Though small dust is clearly present in some disks, the detailed distribution of particle sizes is not well constrained.

Several WDs with dusty disks also possess {\it gaseous} disks \citep{Melis+10}, as inferred by the detection of double-peaked emission lines of Ca II and Fe II (\citealt{Gansicke+06,Gansicke+07,Gansicke+08}).  These features clearly indicate a Keplerian disk composed primarily of gaseous, high-Z elements (no H emission lines have yet been detected) which is radially coincident with the dusty disk \citep{Melis+10}.  Although most DZ WDs display no detectable emission lines, gaseous disks could in principle be more common because the disk line emission is easily overwhelmed by continuum emission from the WD photosphere.  

\citet{Jura03} proposed that compact disks of metals around WDs are produced by the tidal disruption of asteroids or other minor bodies.  A plausible source of such rocky bodies are objects placed onto low periastron orbits by gravitational perturbations from giant planets which have survived the AGB phase of the WD progenitor (\citealt{Duncan&Lissauer98}; \citealt{Debes&Sigurdsson02}).  Disks formed this way are often termed `debris disks' since they result from the destruction of larger bodies, despite the fact that many of their properties (e.g., typical distance from the star, optical depth) are very different from conventional debris disks around young main sequence stars \citep{Wyatt08} as well as the much larger (tens of AU) debris disks around very young, hot ($T\gtrsim 10^5$ K) WDs \citep{Chu+11}.  A tidal disruption origin naturally explains why the outer radii of observed disks $R_{\rm out}\lesssim 1$ R$_\odot$ are similar to the Roche radius $R_R\sim (M_\star/\rho_{d})^{1/3} \sim R_{\odot}$ of a self-gravitating object of typical density $\rho_{d} \sim$ few g cm$^{-3}$, where $M_{\star} \approx 0.6M_{\odot}$ is the WD mass.  

Asteroid-fed disks are also a promising explanation for the metals in the atmospheres of DZ WDs \citep{Jura03}, which otherwise are rapidly depleted by gravitational settling on a timescale much shorter than the WD age (e.g.~\citealt{Paquette+86}).  One can estimate the [appropriately time-averaged] metal accretion rate $\dot{M}_{Z}$ in individual WD systems by combining the observed photospheric metal abundances with theoretical estimates of the mass in the outer convective zone of the WD and the settling times of heavy elements (e.g.~\citealt{Koester&Wilken06}; \citealt{Koester09}).  Values estimated in this way fall within the relatively wide range $\dot M_Z\sim 10^6-10^{10}$ g s$^{-1}$ (\citealt{Farihi+09, Farihi+10a}).  

Several additional lines of evidence now support the asteroid disruption model for WD debris disks and metal pollution.  First, both gaseous and dusty disks are found exclusively around metal-rich WDs: from the statistics of systems observed with {\it Spitzer}, \citet{Farihi+09} estimate that $\sim 20\%$ of DAZ WDs host compact debris disks, yet no evidence is found for disks around WDs without metal pollution (e.g.~\citealt{Hansen+06}; \citealt{Mullally+07}).  \citet{Farihi+10a} also found a strong positive correlation between the presence of a compact debris disk and the metal accretion rate $\dot M_Z$.  The mass in the outer convective zone of many WDs is similar to that of an asteroid of several hundred km size, similar to Ceres or Vesta, while the relative abundances of several elements heavier than He are indeed similar to those in Solar System bodies formed interior to the ice line, such as the Earth, Moon, and asteroids (\citealt{Jura06}; \citealt{Zuckerman+07}; \citealt{Klein+10}; \citealt{Zuckerman+10}).  Finally, the chief alternative theory for WD metal pollution, accretion from the interstellar medium, is now challenged by a variety of observations (e.g.~\citealt{Koester&Wilken06}; \citealt{Kilic&Redfield07}; \citealt{Jura+09a}).

Despite growing evidence in support of the asteroid tidal disruption model, it remains unclear precisely what mechanism(s) are responsible for transporting metals from the disrupted ring of solids (with characteristic radius of several tens of $R_{\star}$) to the WD surface at the observed rates.  The goal of this paper is to address this question with fully self-consistent global numerical models of the coupled evolution of the solid and gaseous components of WD debris disks.    

\begin{figure}
\resizebox{\hsize}{!}{\includegraphics[angle=0]{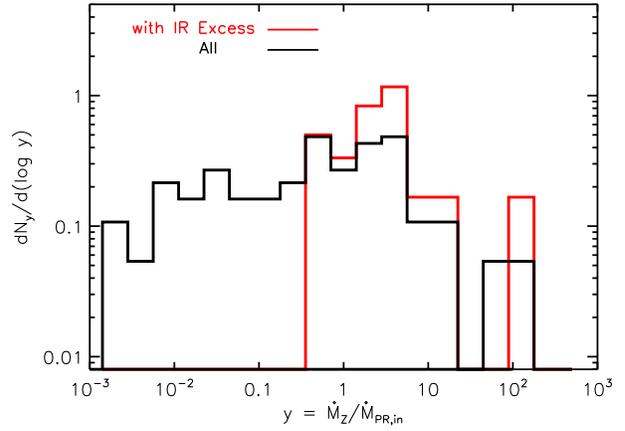}}
\caption[] {Distribution $dN_{y}/d($log $y)$ of WD metal accretion rates $ y \equiv \dot{M}_{Z}/\dot{M}_{\rm PR,in}$ from the samples of \citet{Farihi+09} and \citet{Farihi+10b}, shown separately for the total population (sample size N = 62; {\it black line}) and just those systems with detected IR excess (sample size N = 20; {\it red line}).  We normalize the accretion rate $\dot{M}_{Z}$ to the value $\dot{M}_{\rm PR,in}$ predicted due to optically-thick PR drag (eq.~[\ref{eq:dotM_PR_phys}]), assuming that $R_{\star} \simeq 10^{9}$ cm and using the measured effective temperature of the WD in each system.  Note that each distribution has been separately normalized to unity.  In $\S\ref{sect:applications}$ we show that the current `observed' accretion rates in those systems with $y\gtrsim 1$ and without an IR excess likely overestimate the current real accretion rate (but underestimate the peak accretion rate achieved earlier in their evolution (see Fig.~\ref{fig:mdot_compare}).}
\label{fig:Farihi}
\end{figure}


\subsection{Summary of Previous Work}
\label{sect:past}

We begin by summarizing previous theoretical work on accretion in WD debris disks.  As mentioned above, the inner edge of the solid disk $R_{in}$ is located near the sublimation radius $R_s$ (\citealt{Rafikov11a})
\ba
R_{in} \simeq R_s\equiv \frac{R_\star}{2}\left(\frac{T_\star}{T_s}\right)^{2}\approx 0.2R_{\odot}\left(\frac{R_\star}{0.01R_{\odot}}\right)\left(\frac{T_\star}{10^4\,\mbox{K}}\right)^2\left(\frac{T_{s}}{1500\rm K}\right)^{-2},
\label{eq:R_S}
\ea
where $T_s$ is the sublimation temperature of silicate grains (we assume here that particle's emissivities for starlight and for its own thermal radiation are the same).  Adopting a typical value for the WD radius $R_\star\approx 0.01R_\odot$ \citep{Ehrenreich+11} one finds $R_s\approx 0.2$ R$_\odot$, consistent with observations (\citealt{Jura+07b}; \citealt{Jura+09b}).  

It is worth noting that fitting the SEDs in some WD systems requires an unphysically high sublimation temperature $T_s$ (e.g.~\citealt{Jura+07a}; \citealt{Brinkworth+09}; \citealt{Dufour+10}; \citealt{Melis+10}).  This suggests that other physics, such as a high particle albedo or conduction in the gas phase (\citealt{Jura+07b}), may play an important role, thereby calling the validity of equation (\ref{eq:R_S}) into question.  These complications do not, however, affect the conclusions of this work since hereafter we scale all relevant distances to $R_{in}$ without specifying what physics sets its value.

The sublimation of solid particles produces a source of metallic gas at $r \approx R_s$.  Viscosity redistributes gas from this location, producing an extended disk that transports mass from $R_s$ to the stellar surface $R_{\star}$.  If one assumes that the disk is turbulent, with an effective viscosity characterized by the conventional $\alpha$-parametrization $\nu=\alpha c_s^2/\Omega_K$ \citep{Shakura&Sunyaev73}, then the viscous time at the sublimation radius is given by 
\ba
t_\nu\sim \frac{R_s^2}{\nu}\approx 2\times 10^{3}~\mbox{yr}\left(\frac{\alpha}{10^{-3}}\frac{T_{\rm g}}{5000~\mbox{K}}\right)^{-1}\left(\frac{\mu}{28m_{p}}\right)\left(\frac{M_\star}{0.6 M_\odot}\frac{R_s}{0.2~\mbox{R}_\odot}\right)^{1/2},\nonumber \\ 
\label{eq:t_nu}
\ea
where $c_{s} = (kT_{g}/\mu)^{1/2}$ is the gas sound speed; $T_{g}$ is the gas temperature, normalized to a value characteristic of the range $\approx (3-7)\times 10^{3}$ K set by the balance between photoionization heating and metal line cooling \citep{Melis+10}; $\mu$ is mean molecular weight, normalized to a value $28m_p$ characteristic of pure Si; $\Omega_{K} = (GM_{\star}/r^{3})^{1/2}$ is the Keplerian angular velocity; and $\alpha$ is normalized to a value $10^{-3}$ somewhat lower than that expected to result from MHD turbulence in fully ionized disks (e.g.~\citealt{Davis+10}), an assumption we discuss further in $\S\ref{sect:discussion}$. 

Equation (\ref{eq:t_nu}) shows that the viscous timescale is relatively short compared to other timescales of relevance, such as the total disk lifetime, which is typically estimated to be $\lesssim 10^{5}-10^{6}$ yr (Farihi, private communication).  In most cases the accretion rate onto the stellar surface $\dot M_Z$ is thus ultimately controlled by the rate at which solids reach $R_{\rm s}$, not by the maximum rate of gas accretion set by viscosity.  A key theoretical question is thus how solids are transported to the sublimation radius at rates consistent with the range of values $\dot{M}_{Z} \sim 10^{6}-10^{10}$ g s$^{-1}$ inferred from observations.

One mechanism of solid disk evolution is angular momentum transport due to the interaction between disk particles (e.g.~physical collisions), in full analogy with the rings of Saturn.  However, as we show in Appendix \ref{app:collisional} (see also \citealt{Farihi+08b}), for realistic assumptions about the size distribution and mass of the debris disk, the timescale of this process is generally too long, resulting in a negligible accretion rate $\dot M_Z$.

Another more promising mechanism to drive solids inwards is Poynting-Robertson (PR) drag.  \citet{Rafikov11a} (hereafter R11a) demonstrated that PR drag produces accretion rates up to $\dot M_Z\sim \dot{M}_{\rm PR,in} \lesssim 0.03L_{\star}/c^{2} \sim 10^8$ g s$^{-1}$, where $\dot{M}_{\rm PR,in}$ is the maximum PR accretion rate for an optically thick disk of solids (as defined by eq.~[\ref{eq:sig0}] below) and $L_{\star}$ is the WD luminosity (see eq.~[\ref{eq:dotM_PR}]).  \citet{Bochkarev&Rafikov11} (hereafter BR11) followed this work by computing {\it global} models of the disk evolution under the action of the PR drag.  By considering a variety of initial conditions, including both optically-thin and -thick debris disks, they showed that the results of R11a hold quite generally.  

Figure \ref{fig:Farihi} shows the measured distribution of WD metal accretion rates $\dot{M}_{Z}$ from the samples of \citet{Farihi+09} and \citet{Farihi+10b}.  We normalize $\dot{M}_{Z}$ to $\dot{M}_{\rm PR,in}$ calculated using equation (\ref{eq:dotM_PR}; to be introduced later in $\S\ref{sect:coupled}$), assuming $R_{\rm in} \simeq R_{s}$ (eq.~[\ref{eq:R_S}]) and using the measured effective temperature of each WD in the sample.  Figure \ref{fig:Farihi} shows that a significant fraction of those WD systems with an IR excess (indicative of ongoing metal accretion) have $\dot{M}_{Z} \sim \dot{M}_{\rm PR,in}$ consistent with the characteristic accretion rate due to PR drag (R11a; BR11).  On the other hand, {\it PR drag alone is insufficient to explain the highest observed rates} $\dot M_Z \sim 10^{9}-10^{11}$ g s$^{-1} \gg \dot{M}_{\rm PR,in}$ found in the tail of the distribution in Figure \ref{fig:Farihi}.  

\citet{Rafikov11b} (hereafter R11b) recently proposed a model for the {\it coupled} evolution of the solid and gaseous components of WD accretion disks which can in principle explain even these highest accretion rates.  The key idea is that a fraction of the gas created at the sublimation radius viscously spreads outwards in order to carry away the angular momentum of the accreting gas (\citealt{Pringle81}).  The solid and gaseous disks thus overlap at radii $\gtrsim R_{\rm s}$, providing an natural explanation for their observed radial coincidence (e.g.~\citealt{Melis+10}).  Drag forces between the solid and gaseous disk in this region, resulting from their different rotation rates, act to enhance the rate at which solids lose angular momentum and accrete over the rate provided by PR drag alone.  This scenario results in a feedback process, and potential `runaway', because the enhanced rate of solid sublimation increases the accumulation of gas near the sublimation radius, which in turn enhances the strength of drag and rate of solid accretion.  Using a local model applied to the evolution of the optically thick disk of solids evolution near its inner radius, R11b showed that during such a runaway $\dot M_Z$ can exceed by orders of magnitude the accretion rate $\dot{M}_{\rm PR,in}$ set by PR drag alone.  

R11b has shown that the full range of accretion rates in Figure \ref{fig:Farihi} can in principle be reconciled with the asteroid debris disk model.  However, the R11b model is limited by several simplifying assumptions, including its entirely local (`one zone') nature.  A more detailed, {\it global} model of the gaseous and solid disk evolution is necessary to verify and generalize these conclusions.  


\subsection{This Paper}

In this paper we present one-dimensional calculations of the coupled evolution of the solid and gaseous components of WD debris disks.  Our goal is to test the runaway accretion model of R11b and to begin to address observational tests, such as the expected fraction of metal-bearing WDs with and without observable debris disks and the predicted distribution of accretion rates to compare with data, such as that shown in Figure \ref{fig:Farihi}.  

The paper is organized as follows.  In $\S\ref{sect:model}$ we overview the model, describing in $\S\ref{subsect:aero}$ our baseline description of the aerodynamic coupling between gaseous and solid disks.  In $\S\ref{sect:coupled}$ we present the equations governing the evolution of the coupled gas and solid disks, including an estimate of characteristic values of the dimensionless parameters of the model ($\S\ref{sect:pars}$).  In $\S\ref{sect:results}$ we present our numerical calculations and describe our results.  In $\S\ref{sect:applications}$ we present applications of our results.  These include a comparison between the true accretion rate and that approximated using the commonly adopted assumption of steady-state accretion ($\S\ref{sect:settle}$); and predictions for the accretion rate distribution from an ensemble of debris disk systems ($\S\ref{sect:distribution}$).  In $\S\ref{sect:discussion}$ we discuss our results, including a critical assessment of whether the necessary conditions for runaway are achieved in WD debris disks.  We also discuss several extensions of our standard model, including the effects of an eccentric gaseous disk ($\S\ref{subsect:ecc}$) and the WD magnetic field ($\S\ref{sect:Bfield}$).  In $\S\ref{sect:summary}$ we provide a bulleted summary of our conclusions. 

Appendices of this paper contain a number of important auxiliary results, including an evaluation of the role of collisional viscosity in driving the accretion of solids (Appendix \ref{app:collisional}); calculation of the viscous evolution of a disk with a power law viscosity and mass sources (Appendix \ref{app:1}); an analytical derivation of the runaway condition in a simplified but fully global model of disk evolution (Appendix \ref{app:2}); and a calculation of the aerodynamic coupling between a solid disk and an {\it eccentric} gaseous disk (Appendix \ref{app:ecc}).  Our results in Appendix \ref{app:1} may find useful applications in studies of accretion disks beyond the scope of this work.


\section{Summary of the model}
\label{sect:model}

\begin{figure}
\resizebox{\hsize}{!}{\includegraphics[angle=0]{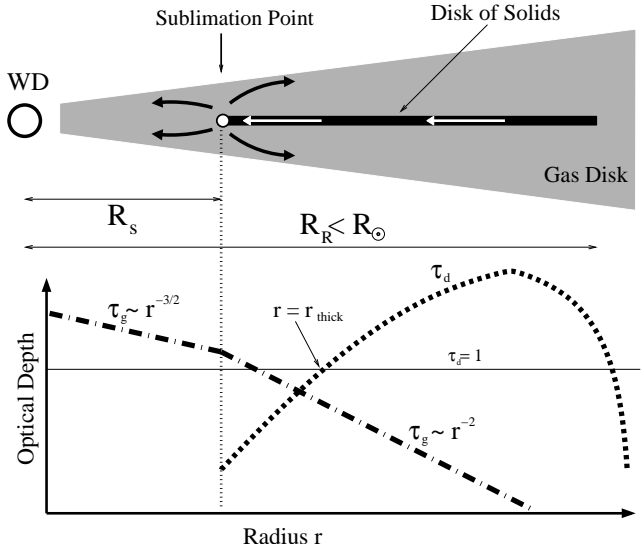}}
\caption[] {Schematic diagram of the model for WD debris disks described in this paper.  An asteroid or other minor body is disrupted by tidal forces inside the Roche radius $R_{R} \sim R_{\odot}$, forming a solid debris disk with surface density distribution $\Sigma_{d}(r) = \Sigma_{0}\tau_{d}(r)$ ({\it dashed line}), where $\tau_{d}$ is the vertical optical depth (eq.~[\ref{eq:sig0}]) as a function of radius $r$.  Poynting-Robertson drag transports solids to the sublimation radius $R_{s} \sim 0.2R_{\odot}$ (eq.~[\ref{eq:R_S}]), where a gaseous disk forms.  Turbulent viscosity redistributes the gas surface density $\Sigma_{g} \propto \tau_{g}$ into a steady-state power-law distribution with a break in the slope (see eq. [\ref{eq:Sig2}]) at the sublimation radius ({\it dot-dashed line}).  The gas and solid disks overlap at radii $\gtrsim R_{s}$, resulting in azimuthal aerodynamic drag between the disks which acts to enhance the rate of solid accretion.  Our calculations in $\S\ref{sect:results}$ show that drag is most effective at the innermost radius $r \sim r_{\rm thick}$ where the solid disk becomes optically thick ($\tau_{d} \gtrsim 1$).  The interaction at this location starts the process which ultimately leads to `runaway' accretion of the entire solid disk.}
\label{fig:schematic}
\end{figure}

We now overview the basic model, which is illustrated schematically in Figure \ref{fig:schematic}.  Initially a solid debris disk forms at radii $r \lesssim R_{\rm R} \sim R_{\odot} \sim 10^{11}$ cm from the tidal disruption of a rocky body.  We characterize the solid debris disk by its surface density $\Sigma_d$ and optical depth
\ba
\tau_d \equiv  \frac{\Sigma_d}{\Sigma_0},~~~~~~~~
\Sigma_0 \equiv  \frac{4}{3}\rho_d a,
\label{eq:sig0}
\ea
where $\Sigma_0$ is the fiducial surface density at which $\tau_d=1$, and $\rho_d \sim$ few g cm$^{-3}$ and $a$ are the bulk density and characteristic size, respectively, of disk particles.  The initial distribution of solids $\tau_{d}(r)$ depends on the details of the tidal disruption process and the properties (e.g., mass, composition, and tensile strength) and incident orbit of the disrupted body (e.g.~\citealt{Debes+12}).

Solids accrete inwards, initially due entirely to PR drag, creating gas at the sublimation radius $R_{\rm in} = R_{s} \approx 2\times 10^{10}$ cm (eq.~[\ref{eq:R_S}]).  The temperature of the gas $T_{g} \sim 3-7\times 10^{3}$ K is higher than that of the solids ($T_{d} < T_{s} \sim 1500$ K) due to differences in their heating and cooling processes \citep{Melis+10}.  The aspect ratio of the gaseous disk $h_{g}/r \sim 10^{-3}$ is typically small, but the disk scale-height $h_{g} \sim 10^{7}$ cm is still much larger than that of the particulate disk, which (with the possible exception of small dust particles) appears to be well settled into the midplane.  

Due to angular momentum transport by turbulent viscosity, the gaseous disk accretes both inwards to the WD surface $R_{\star} \sim 10^{9}$ cm and spreads outwards to radii $\gtrsim \,R_{\rm in}$, where it overlaps with the solid disk.  If the rate of solid accretion evolves slowly compared to the viscous time, then the gas surface density (proportional to optical depth $\tau_{g}$) develops a power-law radial distribution, with a break in the power-law index at the sublimation radius (see Appendix $\ref{app:1}$).  Drag on the solids due to interaction with the gaseous disk enhances the rate of solid accretion.  

Note that in this work we do not consider the possibility of gas production due to particle collisions, as suggested by \citet{Jura08}, \citet{Farihi+09}, since, as we demonstrate in \S\ref{subsect:gas_in_coll}, this mechanism cannot maintain the continuous production of metallic gas on a sufficiently long timescale.

We now discuss one physical mechanism for coupling the dynamics of the solid and gaseous disks: {\it aerodynamic drag}.  As will be shown later in $\S\ref{sect:results}$, aerodynamic drag is most effective at the innermost radius $r \approx r_{\rm thick}$ where the solid disk becomes optically thick ($\tau_{d} \gtrsim 1$).


\subsection{Aerodynamic Coupling}
\label{subsect:aero}

The angular velocity of the gaseous disk $\Omega_g$ is slightly lower than the Keplerian rate $\Omega_K = (GM_{\star}/r^{3})^{1/2}$ at which the solid particles orbit due to the radial pressure gradient in the gaseous disk, $\Omega_g-\Omega_K\approx (2\Omega_K r\rho_g)^{-1}\partial P_g/\partial r$, where $P_g$ and $\rho_g$ are the midplane gas pressure and density, respectively.\footnote{In places hereafter where the distinction between $\Omega_{g}$ and $\Omega_{K}$ is not important, we simply refer to {\it the} angular rotation rate $\Omega = \Omega_{K} \simeq \Omega_{g}$.}   This results in an azimuthal velocity difference given by
\ba
&v_{\varphi,rel}&=-\eta c_s\frac{c_s}{\Omega r}\approx \nonumber \\
& & -200~\mbox{cm s}^{-1}\left(\frac{28m_{p}}{\mu}\right)\left(\frac{T_{\rm g}}{5000~\mbox{K}}\right)\left(\frac{M_\star}{0.6 M_\odot}\frac{R_{in}}{0.2~\mbox{R}_\odot}\right)^{1/2},
\label{eq:v_rel}
\ea
where (assuming $c_s$ independent of radius)
\ba
\eta=\frac{1}{2}\left(\frac{3}{2}-\frac{\partial \ln\Sigma_g}
{\partial \ln r}\right)
\label{eq:eta}
\ea
is a dimensionless parameter set by the radial pressure distribution.  Here $\Sigma_g \simeq 2 h_{g}\rho_g$ and $h_{g} \simeq c_{s}/\Omega$ are the surface density and vertical scale-height of the gas disk, respectively. 

The importance of shear between the solid particles and gaseous disk was first recognized in the context of protoplanetary disks (\citealt{Whipple72}) and is now widely believed to result in a variety of important effects, such as the inward migration of solids \citep{Weidenschilling77}, excitation of turbulence near the disk midplane (e.g.~\citealt{Cuzzi+93}), and streaming instabilities in the disk of solids (\citealt{Youdin&Goodman05}; \citealt{Johansen+07}).  


\subsubsection{Optically Thick Disks ($\tau_d\gtrsim 1$)}
\label{subsect:thick}

One of the most important (and also most uncertain) issues is the strength of the coupling between the gaseous and solid disks.  When the optical depth of the solid disk is high ($\tau_{d} \gg 1$), its geometry may be idealized as that of a solid plate with a zero-slip boundary condition at its surface.  In reality, of course, the disk is not a {\it perfectly smooth} plate; its surface may, for instance, contain structures similar to those observed in the rings of Saturn, such as wakes, vertical particle motions, particle `pileups', and other highly inhomogeneous small-scale surface density features \citep{Cuzzi+10}.  These complications introduce significant uncertainty when calculating the drag force acting on the debris disk. 

Even if the idealization of a smooth plate is adopted, one must determine whether the gaseous layer above the plate is laminar or turbulent.  A similar issue was first addressed by \citet{Goldreich&Ward73}, who showed that if the gaseous disk remains laminar, then the velocity shear between the two disks is restricted to within the Ekman layer of vertical thickness
\ba
z_E\sim\left(\frac{\nu_{m}}{\Omega}\right)^{1/2}\approx
h_{g}\left(\frac{\mu}{\sigma \Sigma_g}\right)^{1/2},
\label{eq:z_E}
\ea
where $\nu_{m}\approx\lambda c_s/3$ is the {\it molecular} shear viscosity in the disk (not to be confused with the {\it turbulent} viscosity $\nu$), $\lambda = \mu/\rho_{g}\sigma$ is the molecular mean free path, and $\sigma$ is the molecular cross section.  The gas surface density near the sublimation radius can be estimated using the expression $\dot{M}_{Z} = 3\pi\nu\Sigma_{g}$ appropriate for steady-state accretion (e.g.~\citealt{Frank+02}; see also eq. [\ref{eq:Sig2}]),
\ba
\Sigma_{g}(r = R_{\rm in}) &\simeq& \frac{\dot{M}_{Z}t_{\nu}}{3\pi R_{s}^{2}} \nonumber \\
&\approx& 3\times 10^{-3}~{\rm g\,cm^{-2}}\left(\frac{\dot{M}_{Z}}{10^{8}\rm \,g\,s^{-1}}\right) \left(\frac{\alpha}{10^{-3}}\right)^{-1}\left(\frac{R_s}{0.2R_{\odot}}\right)^{-3/2}, 
\label{eq:sigma_g}
\ea
where we have assumed $R_{\rm in} \approx R_{\rm s}$ and have substituted equation (\ref{eq:t_nu}) for $t_{\nu} = R_{s}^{2}/\nu$, adopting fiducial parameters of $\mu = 28m_{p}$, $T_{g} = 5000$ K and $M_{\star} = 0.6M_{\odot}$.  

For $\sigma \approx 10^{-15}$ cm$^{-2}$, $\alpha = 10^{-3}$, and the typical range of measured WD accretion rates $\dot{M}_{Z} \sim 10^{8}-10^{10}$ g s$^{-1}$, equation (\ref{eq:z_E}) shows that the Ekman layer thickness $z_E\sim 10^{3}-10^{5}$ cm is generally much smaller than the disk scale-height $h_{g} \sim 10^{7}$ cm.  For the same parameters, the Reynolds number of the Ekman layer is estimated to be 
\ba
Re_E=\frac{v_{\varphi,rel}z_E}{\nu_m} \approx
\frac{h_{g}}{r}\left(\frac{\sigma \Sigma_g}{\mu}\right)^{1/2} \sim 0.1-1.
\label{eq:Re_E}
\ea
Under terrestrial conditions, both numerical calculations \citep{Coleman+90} and experiments \citep{Faller63} find that $Re_E\gtrsim 10^2$ is necessary for the Ekman layer to become turbulent.  At first glance it thus appears that the gas layer above the debris disk is laminar. 

In astrophysical disks, however, the Ekman layer may be susceptible to Kelvin-Helmholtz (KH) instabilities \citep{Goldreich&Ward73}.  If one assumes momentarily that fluid in the shear layer is incompressible ($v_{\varphi,rel}\ll c_s$), then the Richardson number of the Ekman layer is given by
\ba
Ri_E=-g_z\frac{\partial\ln\rho/\partial z}
{(\partial v_\varphi/\partial z)^2}\sim 
\left(\frac{g_z z_E}{c_s v_{\varphi,rel}}\right)^2\approx
\left(\frac{\Omega r}{c_s}\frac{\mu}{\sigma \Sigma_g}\right)^2,
\label{eq:Rich_E}
\ea
where $g_{z} \simeq GM_{\star}z_{E}/r^{3}$ is the vertical gravitational acceleration near the midplane and we have made the approximations $\partial\ln\rho/\partial z\sim g_z/c_s^2$ and $\partial v_\varphi/\partial z\sim v_{\varphi,rel}/z_E$.  Again adopting the fiducial parameters and range of accretion rates given above, one finds $Ri_E\sim 10^{-2}-10^{-6}$, indicating that the flow is grossly KH unstable and must be turbulent.\footnote{A similar conclusion does not apply under terrestrial conditions because $Ri_E \propto g_z^{2}$ and the gravitational acceleration on Earth is $\gtrsim 10^4$ times higher than $g_{z}$ in WD debris disks.}  

If the flow over the disk surface is turbulent, then the force per unit surface area is given by (\citealt{Goldreich&Ward73})
\ba
f_{a}(\tau_d\gtrsim 1)
=-\mbox{Re}_\star^{-1}\rho_g v_{\varphi,rel}|v_{\varphi,rel}|,
\label{eq:aerodrag} \qquad 
\ea
where Re$_\star^{-1}$ is a proportionality constant.  Estimates of the value of Re$_\star$ in the case of a smooth solid plate vary significantly in the literature, ranging from Re$_\star\approx 20$ \citep{Dobrovolskis+99} to Re$_\star\approx 500$ \citep{Goldreich&Ward73}.  Even given this uncertainty, the smooth plate model probably underestimates the true drag, since the surface of a particulate disk is not continuous.  As we describe below, in some cases the gas-solid interaction may be better described as that due to a combination of individual particles, in which case the effective value of Re$_\star$ may be much smaller.


\subsubsection{Optically Thin Disks ($\tau_d\lesssim 1$)}

In the opposite limit of an extremely optically thin disk ($\tau_d\ll 1$), debris particles interact with the gas individually.  The speed of gas relative to solids in this case is just $v_{\varphi,rel}$ (eq.~[\ref{eq:v_rel}]), since the particles do not disturb the mean gas flow appreciably.  The drag force per unit surface area for optically-thin disks is then
\ba
f_a(\tau_d\ll 1)=\frac{\Sigma_d}{(4\pi/3)\rho_{d} a^3}F_D=
\tau_d\frac{F_D}{\pi a^2},
\label{eq:ind}
\ea
where $F_D$ is the drag force on an individual particle of size $a$.  

The magnitude of $F_D$ depends on two key parameters (\citealt{Whipple72}; \citealt{Weidenschilling77}): (1) the ratio of particle size $a$ to the molecular mean free path $\lambda$,
\ba
\frac{a}{\lambda} = \frac{\Sigma_{g}\Omega a}{2\mu c_{s}} \sim 5\,\left(\frac{\dot{M}_{Z}}{10^{8}\rm \,g\,s^{-1}}\right)\left(\frac{\alpha}{10^{-3}}\right)^{-1}\left(\frac{a}{\rm cm}\right);
\label{eq:lambda}
\ea
and (2) the {\it particle} Reynolds number
\ba
Re_p=\frac{2 a v_{\varphi,rel}}{\nu_{m}}\sim\frac{c_s}{\Omega r}
\frac{a}{\lambda}\sim 10^{-2}\left(\frac{\dot{M}_{Z}}{10^{8}\rm \,g\,s^{-1}}\right)\left(\frac{\alpha}{10^{-3}}\right)^{-1}\left(\frac{a}{\rm cm}\right),
\label{eq:Re_p}
\ea
where in equation (\ref{eq:lambda}) we have substituted equation (\ref{eq:sigma_g}) for $\Sigma_{g}$, and in our numerical estimates we again adopt characteristic values for the relevant parameters at $r \sim R_{\rm s}$.  

Equation (\ref{eq:Re_p}) shows that $Re_p \lesssim 1$ is satisfied for $\alpha = 10^{-3}$ and the range of accretion rates $\dot{M}_{Z} \sim 10^{8}-10^{10}$ g cm$^{-3}$ provided that the typical particle size obeys $a \lesssim 1-100$ cm.  Assuming that this condition is indeed satisfied (consistent with observations, \citealt{Graham+90}; see also \S \ref{sect:discussion}), then the interaction of gas with {\it individual particles} is in the laminar regime.  The drag force in this case, in the Stokes ($a \gtrsim \lambda$) and Epstein ($a \lesssim \lambda$) regimes, respectively, is given by the expression 
\begin{equation}
F_{D} =
\left\{
\begin{array}{lr}
6\pi a\rho_g \nu_{m} v_{\varphi,rel}, \qquad &
a \gtrsim \lambda\,\,{\rm(Stokes)}\\
(4\pi/3)\rho_g a^2 c_s v_{\varphi,rel}, \qquad & 
a \lesssim \lambda\,\,{\rm(Epstein)}\\
\end{array}
\right.,
\label{eq:drag}
\end{equation}
These two limits are readily combined into a single formula 
\ba
F_D=-\frac{12\pi a^2}{Re_p}\rho_g v_{\varphi,rel}|v_{\varphi,rel}|\times \mbox{min}
\left(1,\frac{2}{3}\frac{a}{\lambda}\right),
\label{eq:single}
\ea
which substituted into equation (\ref{eq:ind}) gives the force per area
\ba
f_a(\tau_d\ll 1)=-\frac{12}{Re_p}\tau_d\rho_g 
v_{\varphi,rel}|v_{\varphi,rel}|\times \mbox{min}
\left(1,\frac{2}{3}\frac{a}{\lambda}\right).
\label{eq:ind1}
\ea


\subsubsection{Connecting the Optically-Thin and -Thick Regimes}

A comparison of our expressions for the drag force in the optically-thick (eq.~[\ref{eq:aerodrag}]) and optically-thin (eq.~[\ref{eq:ind1}]) regimes shows that they do not match smoothly at $\tau_d\sim 1$, since in general one expects Re$_{\star}^{-1} \ll 1 \ll 12/Re_p$, while $Re_p \propto a/\lambda$ varies independently.  This mismatch is not unexpected, since the transition regime $\tau_d\sim 1$ is precisely where the smooth-plate approximation breaks down.  In reality there is an intermediate range of optical depths, between $\tau_{d} \sim 1$ and some lower value $\tau_{d} \equiv \tau_{d,ind} \ll 1$, between which the solid-gas coupling is not well described as either drag on a {\it continuous} entity, nor as the sum of {\it incoherent} forces on individual particles.  The relative velocity between the gas and particle disk in this regime also lies somewhere between zero (the no-slip boundary condition realized in the case of a smooth plate) and $v_{\varphi,rel}$ (eq.~[\ref{eq:v_rel}]).  

\begin{figure}
\resizebox{\hsize}{!}{\includegraphics[angle=0]{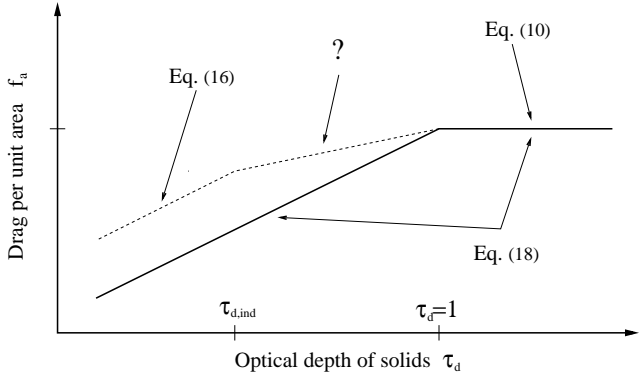}}
\caption[] {Different regimes of aerodynamic drag between the gaseous and particulate disks as a function of the optical depth of solids $\tau_d$.  The drag law used in this work ({\it solid line}; eq.~[\ref{eq:ind1}]) underestimates the drag force $f_a$ in the optically thin limit $\tau_d\lesssim 1$ as compared to the realistic drag law shown by dashed line, although its qualitative behavior is captured.  The true coupling between gas and solids is poorly understood at intermediate values of optical depth $\tau_{d,ind}\lesssim \tau_d\lesssim 1$, at which collective effects influence aerodynamic drag. See text for details.
}
\label{fig:drag_picture}
\end{figure}

In this work we do not model this poorly understood intermediate regime of gas-particle interaction in detail. Instead we adopt the following simple formula for the drag force, which interpolates smoothly between optically thick and optically thin limits:
\ba
f_{a}&=&-\frac{1-e^{-\tau_d}}{\mbox{Re}_\star}\rho_g 
v_{\varphi,rel}|v_{\varphi,rel}|
\label{eq:aerodrag1}\\
&=&\eta|\eta|A_a\Sigma_g\left(1-e^{-\tau_d}\right),
\label{eq:aerodrag2}
\ea
where
\ba
A_a=
{\rm Re}_\star^{-1}\frac{c_s^3}{\Omega r^2},
\label{eq:A_a}
\ea
Although equation (\ref{eq:aerodrag1}) reduces to equation (\ref{eq:aerodrag}) when $\tau_d\gtrsim 1$, it considerably underestimates the drag force for $\tau_d\ll 1$ as compared to equation (\ref{eq:ind1}).  This point is illustrated in Figure \ref{fig:drag_picture}, in which we compare our approximation of the drag force in equation (\ref{eq:aerodrag1}) to a schematic representation of the more realistic drag behavior, which properly reduces to equations (\ref{eq:aerodrag}) and (\ref{eq:ind1}) in the optically thick ($\tau_d\gtrsim 1$) and thin ($\tau_d\lesssim \tau_{d,ind}$) limits, respectively.   Although there are modest differences between these expressions at small values of $\tau_d$, equation (\ref{eq:aerodrag1}) nevertheless does reproduce the qualitative scaling $f_{a}\propto \tau_d$ in the optically thin regime, which is the most important property for the purposes of our current work.  As we describe in $\S\ref{sect:results}$, this distinction between ``weak'' ($\tau_{d} \lesssim 1$) and ``strong'' ($\tau_{d} \gtrsim 1$) coupling has an important role in the build-up to runaway in the accretion disk evolution.

Although our numerical calculations focus on aerodynamic coupling of the above form, in principle other forms of drag may be present, such as enhanced drag due to an eccentric gas disk ($\S\ref{subsect:ecc}$) or electromagnetic inductive coupling with an external magnetic field ($\S\ref{sect:Bfield}$).  To the extent that the drag force remains proportional to the local gas mass, an expression of the form in equation (\ref{eq:aerodrag2}) may be applicable to these cases as well, provided that an appropriate alternative expression is substituted for equation (\ref{eq:A_a}).


\section{Coupled evolution of the particulate and gaseous disks}
\label{sect:coupled}


In this section we present the equations governing the solid and gaseous disks, including their angular momentum exchange due to aerodynamic drag and mass exchange due to sublimation.


\subsection{Solid Disk}  
\label{sect:disk_solid}

The evolution of the solid disk obeys the continuity equation
\ba
\frac{\partial\Sigma_d}{\partial t}-\frac{1}{2\pi r}
\frac{\partial \dot M}{\partial r}=S_d.
\label{eq:cont}
\ea
where $\dot M$ is the mass flux through the solid disk and the function $S_d$ accounts for sources or sinks of solid mass, such as dust sublimation or gas condensation.

The mass flux consists of several components
\ba
\dot M=\dot M_{PR}+\dot M_{a}+\dot M_{d},
\label{eq:dotM}
\ea
which include the mass flux due to PR drag (R11a)
\ba
\dot M_{PR}=\zeta\frac{L_\star}{c^2}\left(1-e^{-\tau_d/\zeta}\right),
\label{eq:dotM_PR}
\ea
where (\citealt{Friedjung85})
\ba
\zeta(r)=\frac{4}{3\pi}\frac{R_\star}{r}
\label{eq:zeta}
\ea
is the incidence angle of the stellar radiation with respect to the flat disk midplane. Note that $\dot M_{PR}$ depends on the optical depth $\tau_d/\zeta$ to stellar light arriving at grazing incidence angle $\zeta\ll 1$.

A second contribution to the mass flux results from aerodynamic drag by the gas disk (R11b):
\ba
\dot M_{a}=\frac{4\pi r f_a}{\Omega}=
\frac{4\pi\eta|\eta| r A_a \Sigma_g}{\Omega}\left(1-e^{-\tau_d}\right),
\label{eq:dotM_a}
\ea
where we have substituted equation (\ref{eq:aerodrag1}) for $f_{a}$ from $\S\ref{subsect:aero}$. The factor $A_a$ in equation (\ref{eq:dotM_a}) depends on the specific form of aerodynamic drag.  In general we assume this function scales as a power law with radius:
\ba
A_a=A_{in}\left(\frac{r}{R_{in}}\right)^\kappa,
\label{eq:A}
\ea
where $\kappa$ is a constant that depends on the drag law.  If $c_{s}$ is constant with radius, our fiducial drag law given in equation (\ref{eq:aerodrag1}) implies $\kappa=-1/2$, a value we therefore adopt throughout the remainder of this paper.

From equation (\ref{eq:dotM_a}) one can define the {\it sublimation} time $t_s$ as the timescale on which the characteristic gaseous mass near the sublimation radius $\pi R_{in}^2\Sigma_g$ is replenished by accretion due to aerodynamic drag (in the optically thick regime $\tau_{d} \gg 1$) alone:
\ba
t_s\equiv \frac{\pi R_{in}^2\Sigma_g}{\dot M_a(r=R_{in},\tau_{d}\gg 1)}=
\frac{\Omega_{in} R_{in}}{4 A_{in}},
\label{eq:t_s}
\ea
where $\Omega_{in}\equiv \Omega(R_{in})$ and in evaluating $\dot{M}_{a}$ (eq.~[\ref{eq:dotM_a}]) we have set $\eta = 1$.  Note that this timescale is independent of $\Sigma_g$.

The final contribution to the mass flux $\dot{M}_{d}$ in equation (\ref{eq:dotM}) accounts for the possible condensation of gas.  This process necessarily exchanges angular momentum between the gas and solid disks, thus driving additional solid accretion.  In this paper we neglect condensation by setting $S_d = \dot M_{d} = 0$ (see $\S\ref{sect:condensation}$).  However, once an expression for $S_d$ is given, the resulting expression for $\dot M_{d}$ is easily calculable.


\subsection{Gaseous Disk}  
\label{sect:disk_gas}

The gaseous disk also evolves according to the continuity equation (\citealt{Lin&Papaloizou96})
\ba
\frac{\partial\Sigma_g}{\partial t}-\frac{1}{r}\frac{\partial}
{\partial r}\left[3r^{1/2}\frac{\partial}
{\partial r}\left(\Sigma_g\nu r^{1/2}\right)-\frac{2\Sigma_g\Lambda}
{\Omega}\right]=S_g,
\label{eq:gas}
\ea
where $\nu$ is again the turbulent viscosity and  
\ba
\Sigma_g\Lambda=\eta|\eta|A_a\Sigma_g r\left(1-e^{-\tau_d}\right).
\label{eq:trq}
\ea
is the torque per unit area on the gaseous disk due to the back-reaction from the aerodynamic drag force $f_a$ (eq.~[\ref{eq:aerodrag1}]) acting on the solid disk. In Appendix \ref{app:2} we explore a global model with back-reaction neglected ($\Lambda=0$) and demonstrate that the main evolutionary features are still preserved even in this simplified setup. 

The function $S_g$ represents sources and sinks of gaseous mass.  The most important source of gas is that due to sublimation, which we assume is sharply localized at $r=R_{in}$ and is proportional to the rate of solid accretion $\dot{M}$:
\ba
S_g=\frac{\dot M(r=R_{in})}{2\pi R_{in}}\delta(r-R_{in}),
\label{eq:gas_source}
\ea
where $\delta(z)$ is the Dirac delta function.

We assume that the gas temperature and sound speed are independent of radius, in which case the turbulent viscosity may be written:
\ba
\nu(r)= \alpha c_{s}h_{g} = \alpha(r)\frac{c_s^2}{\Omega_{in}}
\left(\frac{r}{R_{in}}\right)^{3/2},
\label{eq:nu1}
\ea
where in general we allow the dimensionless $\alpha$-parameter to vary with radius, denoting $\alpha_{in}\equiv \alpha(r=R_{in})$.


\subsection{Dimensionless Equations}  
\label{sect:dim-less}


We now bring the equations from the previous two sections into dimensionless form.  We first introduce several new definitions:
\ba
&& \tilde t\equiv \frac{t}{t_\nu},~~~x\equiv\frac{r}{R_{in}},~~~
\tau_g\equiv\frac{\Sigma_g}{\Sigma_0}, ~~~
\zeta_{in}\equiv \frac{4}{3\pi}\frac{R_\star}{R_{in}},
\label{eq:def2}\\
&& 
s_g\equiv\frac{S_g t_\nu}{\Sigma_0},~~~
\dot M_{PR,in}\equiv\frac{L_\star}{c^2}\zeta_{in},~~~
\tilde \alpha(x)\equiv\frac{\alpha(x)}{\alpha_{in}},
\label{eq:def1}\\
&& \dot m\equiv\frac{\dot M t_s}{\pi R_{in}^2\Sigma_0},~~~
\dot m_d\equiv\frac{\dot M_d t_s}{\pi R_{in}^2\Sigma_0},
\label{eq:def}
\ea
noting in particular that we express time in units of the viscous time $t_{\nu}$ at $R_{\rm in}$ (eq.~[\ref{eq:t_nu}]) and radius in units of the location of the inner edge of the solid disk $R_{in} \simeq R_{s}$ (eq.~[\ref{eq:R_S}]).

We also introduce two key parameters: the {\it feedback parameter}
\ba
{\cal F}\equiv \frac{t_\nu}{t_s}=\frac{4A_{in} R_{in}}{\alpha_{in} c_s^2},
\label{eq:feedback}
\ea
and the {\it Poynting-Robertson (PR) parameter}
\ba
b_{PR}\equiv\frac{\dot M_{PR,in}t_s}{\pi R_{in}^2\Sigma_0}.
\label{eq:PR_par}
\ea
As we will show in $\S\ref{sect:results}$, the feedback parameter has an important qualitative effect on the evolution of WD debris disks, in particular by controlling whether and when runaway accretion occurs.  The physical interpretation of the PR-parameter is less clear, but roughly speaking it represents the ratio of the timescale for disk replenishment due to the gas drag-assisted accretion of solids $t_{s}$ (eq.~[\ref{eq:t_s}]) to the timescale on which the gas disk can be replenished by PR drag alone, given a marginally optically-thick disk (mass $\approx \pi R_{in}^{2}\Sigma_{0}$) concentrated near the sublimation radius.  Note that (1) $b_{PR}$ depends on $\Sigma_0$ (eq.~[\ref{eq:sig0}]) and is thus sensitive to the particle size $a$; (2) from equations (\ref{eq:def}) and (\ref{eq:PR_par}) it follows that $\dot M=\dot M_{PR,in}\dot m/b_{PR}$.

Given these definitions, we now recast the evolutionary equations from $\S\ref{sect:disk_solid}$ and $\S\ref{sect:disk_gas}$ in dimensionless form.  First, the evolution of the solid disk (eq.~[\ref{eq:cont}]) can be written
\ba
\frac{\partial\tau_d}{\partial \tilde t}-\frac{{\cal F}}{2}\frac{1}{x}
\frac{\partial \dot m_{d}}{\partial x}=0,
\label{eq:solid_dim-less}
\ea
where the various contributions to the mass flux of solids are now
\ba
&&\dot m_{d}=\dot m_a+\dot m_{PR},
\label{eq:dotm_dim-less}\\
&& \dot m_a=\eta|\eta|\tau_g x^{(5/2)+\kappa}\left(1-e^{-\tau_d}\right),
\label{eq:dotma_dim-less}\\
&& \dot m_{PR}=b_{PR}x^{-1}\left(1-e^{-\tau_d x/\zeta_{in}}\right),
\label{eq:dotmPR_dim-less}
\ea
with 
\ba
\eta=\frac{1}{2}\left(\frac{3}{2}-\frac{\partial \ln\tau_g}
{\partial \ln x}\right).
\label{eq:eta_dim-less}
\ea

The evolution of the gaseous disk (eq.~[\ref{eq:gas}]) can now be written
\ba
\frac{\partial\tau_g}{\partial \tilde t}-\frac{1}{x}\frac{\partial}
{\partial x}\left[3x^{1/2}\frac{\partial}
{\partial x}\left(\tau_g\tilde\alpha x^2\right)-\frac{{\cal F}}{2}
\dot m_a\right]=s_g,
\label{eq:gas_dim-less}
\ea
with the sublimation source term (eq.~[\ref{eq:gas_source}]) 
\ba
s_g=\frac{{\cal F}}{2}\dot m_{d}(x=1)\delta(x-1).
\label{eq:s_g_dim-less}
\ea

The above expressions represents a system of two equations (\ref{eq:solid_dim-less}) and (\ref{eq:gas_dim-less}) for 2 unknown functions $\tau_d$ and $\tau_g$ and $x$ and $\tilde{t}$, with 3 independent parameters --- ${\cal F}$, $b_{PR}$, and $\zeta_{in}$ --- and 2 functions of $x$ that must be independently supplied --- $\tilde \alpha(x)$ and $A(x)$ (or $\kappa$).


\subsection{Characteristic Values of Key Parameters}  
\label{sect:pars}


We now estimate characteristic values of the key parameters in our model, which are useful for scaling our solutions in $\S\ref{sect:results}$ to arbitrary physical situations.  

First, we estimate the characteristic accretion rate set by PR drag under optically-thick conditions (eq.~[\ref{eq:def1}])
\ba
\dot{M}_{\rm PR,in} = \frac{L_{\star}}{c^{2}}\zeta_{in} \approx 7\times 10^{7}{\,\rm g\,s^{-1}}\left(\frac{R_{\star}}{0.01R_{\odot}}\frac{T_{\star}}{10^{4}\,\rm K}\frac{T_s}{1500\,\rm K}\right)^{2},
\label{eq:dotM_PR_phys}
\ea 
where $\zeta_{in} = 4R_{\star}/3\pi R_{\rm in}$ (eq.~[\ref{eq:def2}]), $L_{\star} = 4\pi R_{\star}^{2}\sigma_{sb}T_{\star}^{4}$ and we have again assumed $R_{\rm in} = R_{\rm s}$.

Using our fiducial model for aerodynamic drag given in equation (\ref{eq:A_a}), the feedback parameter (eq.~[\ref{eq:feedback}]) can be written
\ba
{\cal F} &=& \frac{4}{{\rm Re}_\star\alpha_{in}}
\frac{c_s}{\Omega_{in} R_{in}} \label{eq:F_a} \nonumber \\
&\approx & 7{\rm Re_{\star}^{-1}}\left(\frac{\alpha_{in}}{10^{-3}}\right)^{-1}\left(\frac{T_{\rm g}}{5000\,\rm K}\right)^{1/2}\left(\frac{M_{\star}}{0.6M_{\odot}}\right)^{-1/2}\left(\frac{R_{\star}}{0.2R_{\odot}}\right)^{1/2}
\label{eq:Fnum}
\ea

Depending on the uncertain value of Re$_{\star} \sim 1-100$ (see discussion in $\S\ref{subsect:thick}$), equation (\ref{eq:Fnum}) shows that strong feedback ${\cal F}\gtrsim 1$ requires a rather low value of the viscosity $\alpha_{in} \lesssim 10^{-2}-10^{-4}$ at the sublimation radius.  We will discuss the value of $\mathcal{F}$ in actual debris disk systems further in $\S\ref{sect:discussion}$.

Finally, we estimate characteristic values of the PR parameter $b_{PR}$ (eq.~[\ref{eq:PR_par}])
\ba b_{PR} &= & \frac{4}{\pi}\mathcal{F}^{-1}\frac{\sigma_{sb}(GM_{\star})^{1/2}}{c^{2}\rho_{d}}\frac{R_{\star}^{3}}{R_{\rm in}^{5/2}}\frac{T_{\star}^{4}}{\alpha c_{s}^{2}a} \nonumber \\
&\approx & 3\times 10^{-3}\mathcal{F}^{-1}\left(\frac{\alpha_{\rm in}}{10^{-3}}\right)^{-1}\left(\frac{a}{\rm cm}\right)^{-1}\left(\frac{R_{\rm in}}{0.2R_{\odot}}\right)^{-5/2}\left(\frac{T_{\star}}{10^{4}\,\rm K}\right)^{4}\times \nonumber \\
&&\left(\frac{R_{\star}}{0.01R_{\odot}}\right)^{3}\left(\frac{M_{\star}}{0.6M_{\odot}}\right)^{1/2}\left(\frac{\mu}{28m_{p}}\right)\left(\frac{\rho_{d}}{2.5\,\rm g\,cm^{-3}}\right)^{-1},
\label{eq:bPR}
\ea
where we have used the definitions for $\dot M_{\rm PR,in}$ (eq.~[\ref{eq:def1}]), $\Sigma_0$ (eq.~[\ref{eq:sig0}]), $\mathcal{F}$ (eq.~[\ref{eq:feedback}]), and $R_s$ (eq.~[\ref{eq:R_S}]).  Note that the above estimate is independent of the specific model for aerodynamic drag, except implicitly through the value of ${\cal F}$.  Although $b_{PR} \propto R_{in}^{3}T_{\star}^{4}$ appears to depend sensitively on the WD effective temperature and the inner radius of the solid disk, if the latter coincides with the sublimation radius $R_{\rm in} \approx R_{s} \propto R_{\star}T_{\star}^{2}$ (eq.~[\ref{eq:R_S}]) then the dependence becomes less sensitive, $b_{PR} \propto R_{\star}^{1/2}T_{\star}^{-1}$.

Equation (\ref{eq:bPR}) shows that for typical ranges in the values of the most uncertain parameters $a \sim 0.1-10$ cm and $\alpha_{in} \sim 10^{-4}-10^{-3}$ under conditions of positive feedback $\mathcal{F} > 1$, one finds relatively small characteristic values $b_{PR} \sim 10^{-4}-1$.  Note that a low value of $b_{PR}$ increases $\dot M/\dot M_{PR,in}$ for a given $\dot m$ (see the text after eq.~[\ref{eq:PR_par}]).


\section{Numerical Calculations}
\label{sect:results}

In this section we present our numerical calculations of WD debris disk evolution.

\subsection{Technical Preliminaries}
We evolve the gas surface density $\tau_{g}$ (eq.~[\ref{eq:gas_dim-less}]) using the 2N-RK3 scheme described in \citet{Brandenburg03}.  The solid surface density $\tau_{d}$ (eq.~[\ref{eq:solid_dim-less}]) is evolved as described in BR11.  The gas and solid evolution are coupled at each timestep by the aerodynamic torque $\propto \dot{m}_{a}$ (eq.~[\ref{eq:dotma_dim-less}]) and the point source of gas due to sublimation $\propto s_{g}$ (eq.~[\ref{eq:s_g_dim-less}]).  We adopt a logarithmic radial grid that extends from a location near the WD surface ($x \lesssim 0.1$), out to a radius that at all times safely exceeds the outer edge of the solid distribution ($x \gtrsim 10$).

The initial gas density is set equal to zero at all radii, $\tau_{g}(x,t=0) = 0$, while different initial distributions (e.g., ``gaussian ring'' and ``top-hat'') are adopted for the dust surface density, as described below.  The inner boundary condition on $\tau_{g}$ is set to enforce a constant mass accretion rate $\propto \Sigma_{g}\nu$ across the ghost zones using the value in the first active zone, while $\tau_{g}$ is interpolated across the outer boundary.  We confirm that our results are insensitive to the precise boundary conditions adopted.  We have also verified that the code conserves total mass and angular momentum by checking that any decreases in their values with time are compensated by their fluxes across the inner grid cell (see \citealt{Metzger+09}).

\subsection{Optically Thin Ring}
\label{sec:thinring}

\begin{figure}
\resizebox{\hsize}{!}{\includegraphics[angle=0]{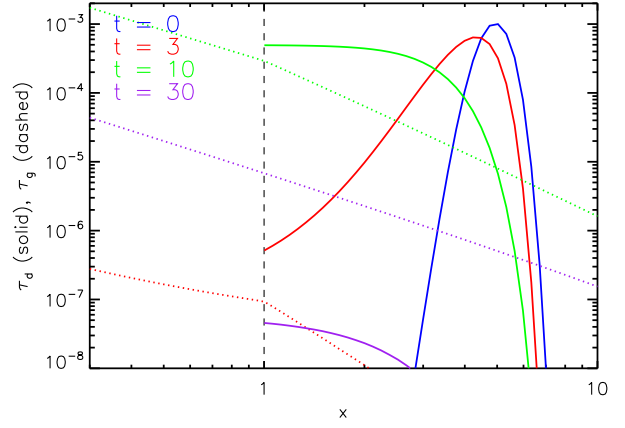}}
\caption[] {Evolution of the surface density profiles of solids ($\tau_{d}$; {\it solid lines}) and gas ($\tau_{g}$; {\it dotted lines}), assuming an initially optically thin ring of solids ($\tau_{d,0} = 10^{-3}$) concentrated at $x_{0} = 5$ (eq.~[\ref{eq:taud_ring}]).  Snapshots are shown at several times $t = 0$ ({\it blue}), $t = 3$ ({\it red}), $t = 10$ ({\it green}), and $t = 30$ ({\it purple}) in units of the viscous time $t_{\nu}$ at the sublimation radius $x = 1$ ({\it vertical dashed line}).  The calculation assumes typical values for the parameters $b_{\rm PR} = 10^{-2}$, $\zeta_{\rm in} = 0.04$, and $\mathcal{F} = 10$.  Note that the evolution of $\tau_{d}$ is very similar to the case without gas drag studied by BR11, even though the level of feedback is high ($\mathcal{F} \gg 1$).  At times when the sublimation rate evolves slowly, the evolution of $\tau_{g}$ is well described by the broken power-law, as predicted by the analytic steady-state solution to the viscous diffusion equation in the presence of a point source of gas at $x = 1$ (see Appendix \ref{app:1}).}
\label{fig:tauthin}
\end{figure}

\begin{figure}
\resizebox{\hsize}{!}{\includegraphics[angle=0]{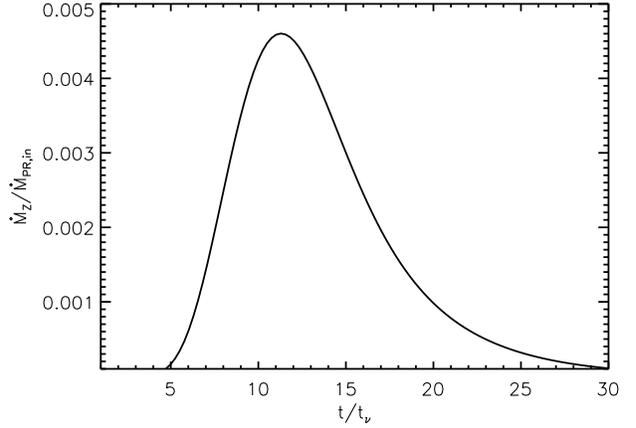}}
\caption[] {Gas accretion rate onto the WD surface $\dot{M}_{Z} \equiv \dot{M}(x = 0)$ in units of $\dot{M}_{\rm PR,in}$ (eq.~[\ref{eq:def1}]) as a function of time for the calculation shown in Figure \ref{fig:tauthin}.  The accretion rate peaks on a timescale $t_{\rm peak} \approx 10$ $t_{\nu}$ that is well approximated by the analytic estimate in equation (\ref{eq:t_thin}).  }
\label{fig:mdotthin}
\end{figure}

We begin by considering the evolution of a thin ring of solid debris with an initial radial profile of the form
\begin{eqnarray}
\tau_{d}(x,t = 0) = \tau_{d,0}\exp\left[-\frac{(x-x_{0})^{2}}{(\Delta x)^{2}}\right],
\label{eq:taud_ring}
\end{eqnarray}
where $x_{0} > 1$ and $\Delta x$ specify the ring center and its radial thickness, respectively.  Although the distribution of solid debris following the tidal disruption of an asteroid may not be well described by a gaussian ring, we focus initially on this simple distribution because it allows us to explore the dependence of the disk evolution on the characteristic optical depth and location of the solid material.  In physical units the total mass in the initial disk is given by
\ba
M_{tot} &=& \int_{R_{\rm in}}^{\infty}2\pi r \Sigma_{d}dr \approx \pi(x_{0}R_{\rm in})^{2}(\Delta x/x_{0})\Sigma_{0}\tau_{d,0} \nonumber \\
&\approx& 10^{21}{\,\rm g}\,\,\tau_{d,0}\left(\frac{\Delta x}{x_{0}}\right)\left(\frac{a}{\rm cm}\right)\left(\frac{x_{0}}{5}\right)^{2},
\label{eq:Mtot}
\ea
where we have used the definition $\Sigma_{0} = (4/3)\rho_{d}a$ (eq.~[\ref{eq:sig0}]) and have assumed characteristic values for $R_{s} = 0.2R_{\odot}$ and $\rho_{d} = 2.5$ g cm$^{-3}$.

Figure \ref{fig:tauthin} shows our calculation of the evolution of an {\it optically thin} ring with $\tau_{d,0} = 10^{-3}$, $x_{0} = 5$ and $\Delta x = 0.5$, corresponding to $M_{tot} = 10^{17}$ g for $a = 1$ cm.  Note that in this example the solid disk is optically thin both in the vertical direction ($\tau_{d} \ll 1$) and horizontally to the incident stellar radiation, for which the condition $\tau_{\Vert} \equiv \tau_{d}/\zeta \ll 1$ is instead relevant (see eq.~[\ref{eq:dotM_PR}] and surrounding discussion).  We adopt typical values for the parameters $b_{PR} = 10^{-2}$ (eq.~[\ref{eq:bPR}]), $\zeta_{in} = 0.04$ (eq.~[\ref{eq:zeta}]), and $\mathcal{F} = 10$ (eq.~[\ref{eq:Fnum}]), the latter allowing for the possibility of strong gas-solid feedback.  Solid and dashed lines show the radial distribution of $\tau_{d}$ and $\tau_{g}$, respectively, at several times $t = 0,3,10,30$ (in units of the viscous time $t_{\nu}$ at $x = 1$) with different colors.  Figure \ref{fig:mdotthin} shows the gas accretion rate at the inner boundary $\dot{M}_{Z} \equiv \dot{M}(x=0)$ in units of the optically-thick PR rate $\dot{M}_{PR,in}$ (eq.~[\ref{eq:def1}]).   

Figures \ref{fig:tauthin} and \ref{fig:mdotthin} illustrate that the evolution of $\tau_{d}$ is very similar to the case without gas drag, as explored previously by BR11.  BR11 show that when the disk is optically thin with $\tau_{d}/\zeta < 1$ at all radii (as satisfied in this case), the dust at radius $x$ accretes on a characteristic timescale set by PR drag
\begin{eqnarray}
t_{\rm acc,thin} \approx \frac{\pi r^{2}\Sigma_{d}}{\dot{M}_{\rm PR}(\tau_{d}/\zeta \ll 1)} = \frac{\pi r^{2}\Sigma_{0}}{\dot{M}_{\rm PR,in}\zeta_{\rm in}^{-1}} = \frac{\zeta_{in}x^{2}}{b_{\rm PR}\mathcal{F}}t_{\nu},
\label{eq:t_thin}
\end{eqnarray}
where $\dot{M}_{\rm PR}(\tau_{d}/\zeta \ll 1) \simeq L_{\star}\tau_{d}/c^{2} = \dot{M}_{\rm PR,in}\tau_{d}\zeta_{in}^{-1}$ is the accretion rate due to PR drag in the optically-thin limit (eq.~[\ref{eq:dotM_PR}]) and we have used the definitions in equations (\ref{eq:PR_par}), (\ref{eq:zeta}) and (\ref{eq:feedback}) for the dimensionless parameters $b_{PR}$, $\zeta_{in}$, and $\mathcal{F}$, respectively.  Note that although we scale $t_{\rm acc,thin}$ to the viscous time in equation (\ref{eq:t_thin}) to compare most easily with our numerical results, we emphasize that the accretion time in the optically thin case depends solely on the rate of PR drag when $t_{\rm acc,thin} \gtrsim t_{\nu}$.

As time advances, the solid ring thus drifts inwards and broadens, the latter simply because the accretion timescale $t_{\rm acc,thin}$ decreases $\propto x^{2}$, consistent with the evolution of $\tau_{d}$ in Figure \ref{fig:tauthin}.  Substituting the relevant parameters, equation (\ref{eq:t_thin}) predicts that the most of the solids reach $x=1$ on a timescale $t_{\rm acc,thin}(x = x_{0}) \approx 10$ $t_{\nu}$ which is set by the initial peak of the dust distribution $x \approx x_{0} = 5$.  This estimate agrees well with the timescale that the gas accretion peaks in our solution (Fig.~\ref{fig:mdotthin}), as is expected because gas accretion rate at the origin faithfully tracks the rate of solid sublimation when the latter varies on timescales greater than the viscous time $t_{\nu}$.  

We conclude that gas coupling does not significantly affect the evolution of optically-thin debris disks.  We have confirmed this by running an otherwise identical calculation with the gas coupling turned off, i.e.~artificially setting the term $\dot{M}_{a} = 0$.  We find a negligible difference between the solutions with and without gas drag, even though `feedback' is nominally strong ($\mathcal{F} \gg 1$).     

Figure \ref{fig:tauthin} also shows the evolution of the gas surface density $\tau_{g}$.  At times when the rates of gas sublimation and accretion evolve relatively slowly, the density profile is well described by a broken power-law distribution $\tau_{g} \propto x^{-n}$ (for $x < 1$) and $\tau_{g} \propto x^{-(n+1/2)}$ ($x > 1$) for $n = 3/2$.  As shown in Appendix \ref{app:1}, this profile is consistent with steady-state analytic solutions of the diffusion equation (eq.~[\ref{eq:gas}] with $\Lambda = 0$) given a viscosity of the form $\nu \propto r^{n}$ ($n = 3/2$ for an isothermal $\alpha$-disk model assumed here) and a constant source of mass at $x = 1$ (see eq.~[\ref{eq:Sig2}]).  Solid-gas coupling thus also has little impact on the gas distribution in optically-thin debris disks.


\subsection{Optically Thick Ring} 
\label{sec:thickring}

\begin{figure}
\centering
\subfigure[With Gas Drag]{
\includegraphics[width = 0.48\textwidth]{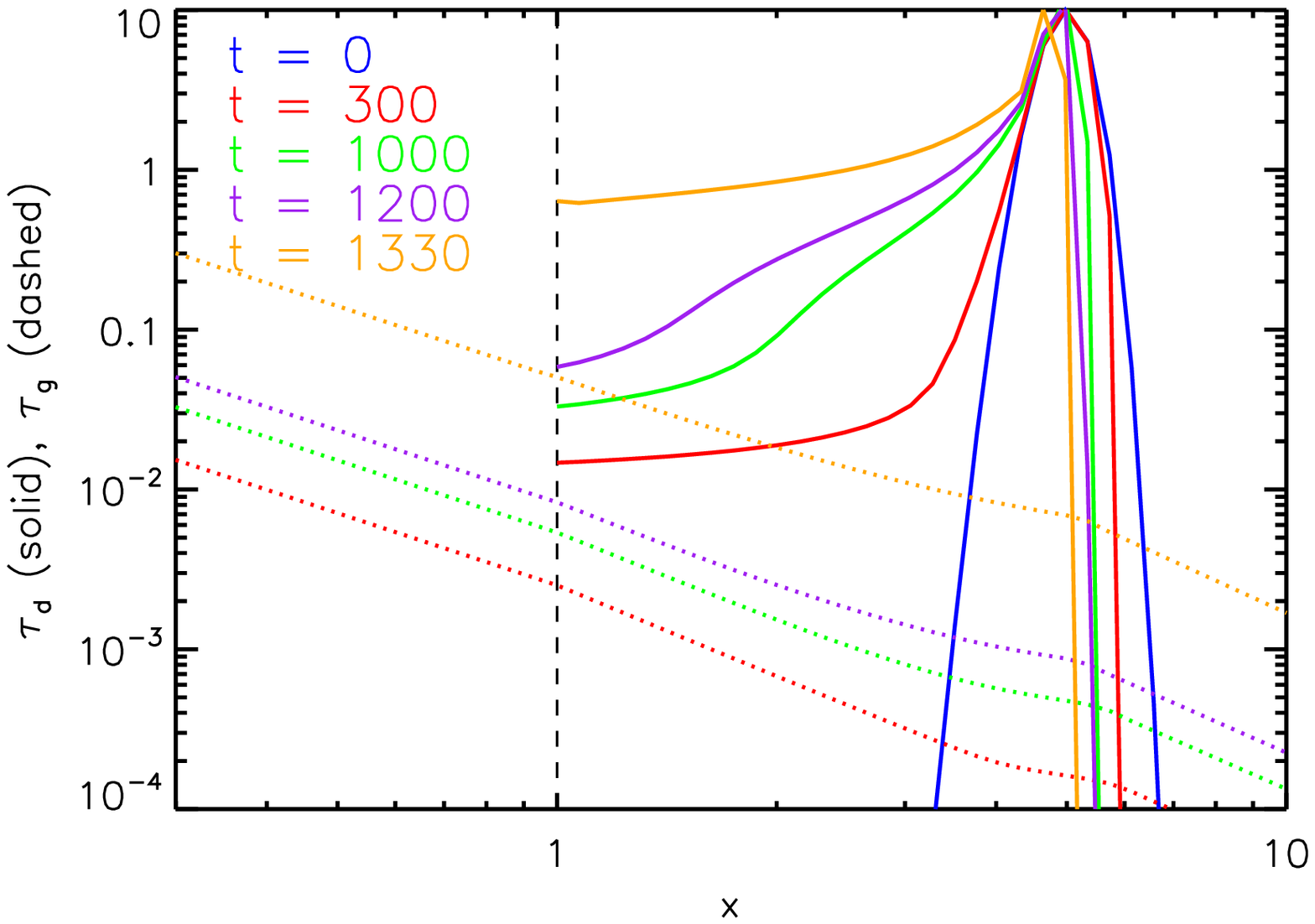}}
\subfigure[Without Gas Drag]{
\includegraphics[width = 0.48\textwidth]{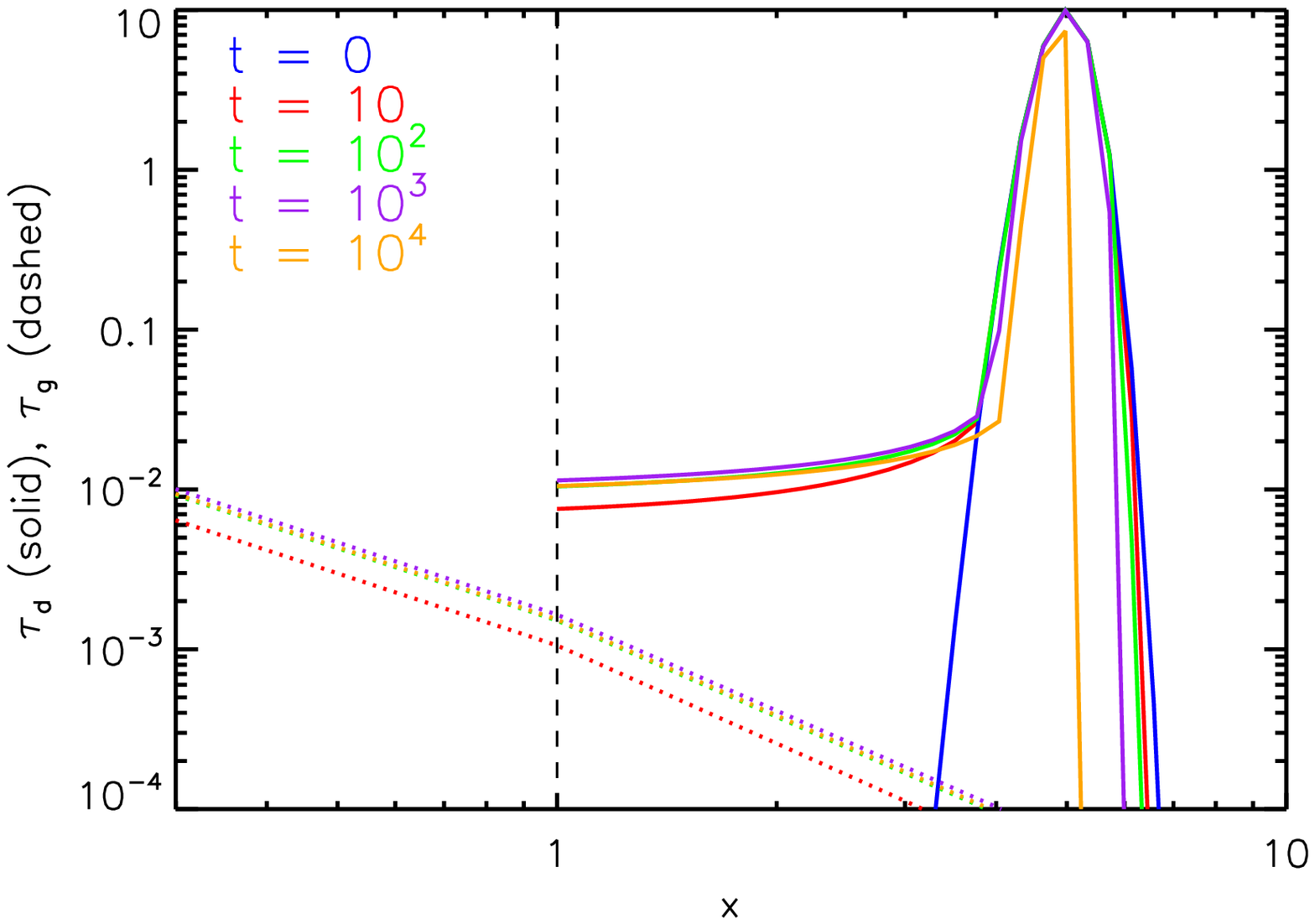}}
\caption[]{Similar to Figure \ref{fig:tauthin}, but now calculated for an {\it optically thick} ring of dust with $\tau_{d,0} = 10$, $x_{0} = 5$, and $\Delta x = 0.5$.  Snapshots are shown at several times with different colors, in units of the viscous time at $x = 1$.  The calculation is performed for parameters $b_{PR} = 10^{-2}$, $\zeta_{in} = 0.04$, and $\mathcal{F} = 2$.  The top panel (a) shows the calculation including aerodynamic drag between the gas and solids $\dot{M}_{a}$, while the bottom panel (b) shows the same calculation but with the drag artificially turned off.}
\label{fig:tauthick}
\end{figure}

\begin{figure}
\resizebox{\hsize}{!}{\includegraphics[angle=0]{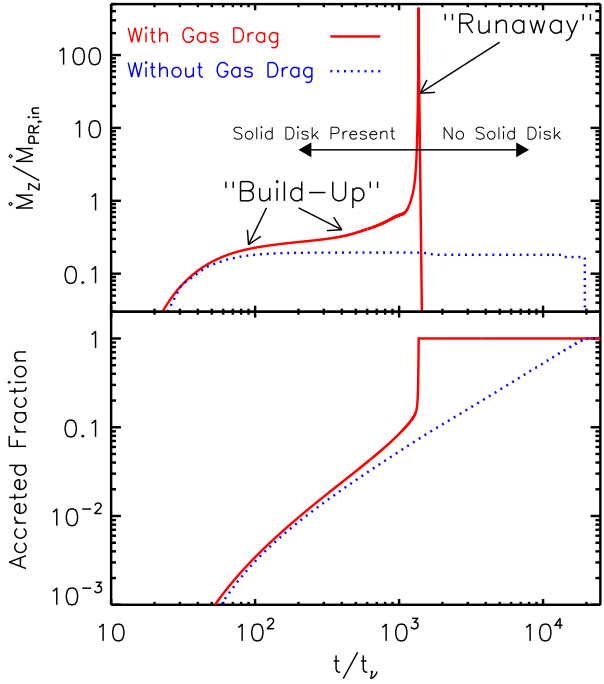}}
\caption[] {({\it Top Panel}) Gas accretion rate onto the WD surface $\dot{M}_{Z}$ in units of $\dot{M}_{\rm PR,in}$ as a function of time for the calculation shown in Figure \ref{fig:tauthick}.  Note that aerodynamic drag, starting at the inner edge of the optically thick part of the disk and moving inwards with time, causes an exponential `build-up' and eventual `runaway' in the accretion rate on a timescale $t \approx 10^{3}t_{\nu}$. ({\it Bottom Panel}) Fraction of the accreted mass for the calculation shown in the top panel.  Note that most of the total mass is accreted during the final runaway phase.}
\label{fig:mdotthick}
\end{figure}

Although gas drag has no significant effect on the evolution of optically-thin disks, it can be substantially more important for optically-thick disks.  Figure \ref{fig:tauthick} shows our calculation of the disk evolution in the case of an initially optically-thick ring of dust ($\tau_{d,0} = 10$), again calculated for characteristic parameters $b_{\rm PR} = 10^{-2}$, $\zeta_{in} = 0.04$, $x_{0} = 5$, $\Delta x = 0.5$, but now for only marginally strong feedback $\mathcal{F} = 2$.  The top panel shows the full calculation including gas drag in the same format as Figure \ref{fig:tauthin}, while for comparison the bottom panel shows for the same calculation with the drag artificially turned off.  Figure \ref{fig:mdotthick} shows the gas accretion rate onto the WD surface $\dot{M}_{Z} \equiv \dot{M}_{g}(x=0)$ as a function of time for both cases. 

First note from Figure \ref{fig:tauthick} that, in cases both with and without gas drag, the disk develops an extremely sharp profile near its outer edge.  This sharp cut-off results from the exponential saturation of the accretion rates both due to PR drag $\dot{M}_{PR} \propto 1-\exp[-\tau_{d}/\zeta]$ (eq.~[\ref{eq:dotM_PR}]) and due to aerodynamic drag $\dot{M}_{a} \propto 1-\exp[-\tau_{d}]$ (eq.~[\ref{eq:dotM_a}]), which are significantly more effective per unit mass in optically-thin regions of the disk.  Since the outer edge of the initial solid distribution is necessarily optically thin, this results in a pile-up of material at the location where the outer edge becomes moderately optically thick (BR11).    

Figures \ref{fig:tauthick} and \ref{fig:mdotthick} also illustrate the dramatic effect that gas drag has on the disk evolution.  Again, in the cases both with and without drag, the distribution of solids initially develops an optically-thin `tail' ($\tau_{d} \ll 1$) at radii between the sublimation point $x = 1$ and the innermost location where the disk first becomes optically thick $x = x_{\rm thick} \lesssim x_{0}$.  

Without gas drag (panel b) the tail supplies an accretion rate from the optically thick disk to the sublimation radius which is approximately constant in time and radius. As shown in BR11, its value $\dot{M}_{\rm PR}(x \approx x_{\rm thick}; \tau_{d} \gtrsim 1) \approx \dot{M}_{\rm PR,in}/x_{0}$ is set by the rate due to PR drag from the inner edge of the optically-thick ring.  The timescale for the {\it entire} ring of dust to accrete through the tail is thus given by
\begin{eqnarray}
t_{\rm acc,thick} \simeq \frac{M_{tot}}{\dot{M}_{\rm PR,in}/x_{0}} \approx \frac{x_{0}^{3}\tau_{d,0}(\Delta x/x_{0})}{b_{\rm PR}\mathcal{F}}t_{\nu}\qquad{\rm (No\,Gas\,Drag)},
\label{eq:t_thick_decoupled}
\end{eqnarray}
where $M_{\rm tot}$ is the total disk mass from equation (\ref{eq:Mtot}) and we have used the definition of $b_{PR}$ from equation (\ref{eq:PR_par}).  Substituting the relevant parameters for the above calculation ($x_{0} = 5$; $b_{\rm PR} = 10^{-2}$; $\tau_{d,0} = 10$; $\Delta x = 0.5$) one finds a long accretion timescale $t_{\rm thick} \approx 10^{4}$ $t_{\nu}$ and a relatively low accretion rate $\dot{M}_{Z} \approx 0.2\dot{M}_{\rm PR,in}$, both consistent with the results of the ``Without Gas Drag'' case shown in the top panel of Figure ~\ref{fig:mdotthick}.  

With gas drag, however, the evolution of $\dot{M}_{Z}$ is dramatically altered (top panel).  Instead of the accretion rate remaining approximately constant in time, Figure \ref{fig:mdotthick} shows that $\dot{M}_{Z}$ initially increases exponentially on a timescale $t \approx 10^{3}t_{\nu}$ (a `build up' stage), before `running away' at a super-exponential rate.  During this process the entire solid disk is consumed on a timescale which is a factor $\sim 10$ times shorter than in the case without gas drag, with the accretion rate during the final runaway reaching a peak value $\sim 10^{3}$ times higher than the PR rate.  As discussed in $\S\ref{sect:past}$, such a runaway process was predicted by R11b, who showed that if feedback is strong ($\mathcal{F} \gtrsim 1$; $t_{\nu} \gtrsim t_{s}$) then the surface density of gas builds up near the sublimation radius $x \gtrsim 1$ due to drag-induced accretion faster than it can be reduced by viscous diffusion.  Because the drag-induced accretion rate $\dot{M}_{a}$ is itself proportional to the gas surface density (eq.~[\ref{eq:dotM_a}]), the build-up of mass at radii $x \gtrsim 1$ results in a runaway.  

Our results confirm that a runaway occurs, but the details of the process differ somewhat from those predicted by R11b.  The simplified `one zone' model of R11b assumed that the solid disk at radii $x \gtrsim 1$ is at all times optically thick.  When $\tau_{d} \gg 1$ the rate of gas drag accretion $\dot{M}_{a} \propto 1-\exp(-\tau_{d})$ (eq.~[\ref{eq:dotM_a}]) saturates at the maximum value, such that runaway grows on the sublimation timescale $\approx t_{s} = t_{\nu}/\mathcal{F}$ (eq.~[\ref{eq:t_s}]).  In our calculation, however, Figure \ref{fig:mdotthick} shows that the runaway is delayed until a timescale $t \equiv t_{\rm run} \sim 10^{3}t_{\nu}$ which is much longer than $t_{s}$.  As we now discuss, this delay occurs because the solids at radii $x \gtrsim 1$ are initially part of the {\it optically-thin} tail created by PR-drag, which couples relatively weakly to the gas.  

Although gas drag is weak just outside the sublimation radius where $\tau_{d} \ll 1$, the gas distribution extends to radii $x \gg 1$ due to viscous spreading (see eq.~[\ref{eq:sig_gas}] below), where it overlaps with the optically-thick solid disk.  Figure \ref{fig:tauthick} shows that the drag force between this gas tail and solids at the inner edge of the optically-thick disk $x \approx x_{\rm thick}$ creates an ``enhancement'' in the solid profile with $\tau_{d} \sim 1$.  The `build-up' stage occurs as this enhancement moves inwards with time.  Finally, once the enhancement reaches the sublimation radius, the condition $\tau_{d}(r \sim R_{s}) \approx 1$ is satisfied and the full runaway occurs in much the same way as predicted by R11b, albeit after a much longer delay time $t_{\rm run} \gg t_{s}$.      

The delay until runaway $t_{\rm run}$ can be estimated as the time required for the mass associated with $\tau_{d}\approx 1$ to propagate from $x \approx x_{\rm thick} \lesssim x_{0}$ to $x = 1$ at the rate set by gas drag accretion $\dot{M}_{a}$:
\begin{eqnarray}
t_{\rm run} \simeq \frac{\pi x_{\rm thick}^{2}\Sigma_{0}}{\dot{M}_{a}(x \approx x_{\rm thick};\,\tau_{d}\approx 1)} \approx \frac{x_{\rm thick}^{3}}{b_{\rm PR}\mathcal{F}^{2}}t_{\nu},
\label{eq:t_run}
\end{eqnarray}
where we have used the definitions of $A_{in}$, $b_{PR}$, $\mathcal{F}$ from equations (\ref{eq:A}), (\ref{eq:PR_par}), and (\ref{eq:feedback}), respectively.  We calculate $\dot{M}_{a}(x \approx x_{\rm thick})$ (eq.~[\ref{eq:dotM_a}]) using the analytic expression for the gas surface density 
\begin{eqnarray}
\Sigma_{g}(r) = \frac{\dot{M}_{d}(x = 1)t_{\nu}}{3\pi r^{2}}
\label{eq:sig_gas}
\end{eqnarray}
derived in Appendix \ref{app:1} (eq.~[\ref{eq:Sig2}]) assuming $\nu \propto r^{n}$ for $n = 1$ (eq.~[\ref{eq:nu1}]) and approximating the sublimation rate by its initial PR-driven value $\dot{M}_{d}(x = 1) \simeq \dot{M}_{\rm PR,in}/x_{0}$.  We also adopt $\eta = 7/4$ appropriate for $\tau_{g} \propto r^{-2}$ (eq.~[\ref{eq:eta}]).   

In order to verify that the above explanation is self-consistent, one must check that the accretion rate due to aerodynamic drag indeed exceeds that due to PR drag at the inner edge of the optically thick disk.  Their ratio is given by
\ba
\left.\frac{\dot{M}_{a}}{\dot{M}_{PR}}\right|_{x \approx x_{\rm thick}; \tau_{d}\approx 1} \approx \frac{49}{48}\mathcal{F},
\label{eq:mdot_ratio}
\ea
where we have again assumed $\eta = 7/4$ and made use of equations (\ref{eq:A}), (\ref{eq:feedback}), and (\ref{eq:dotM_a}) for $A_{in}$, $\mathcal{F}$, and $\dot{M}_{a}$, respectively.  Equation (\ref{eq:mdot_ratio}) confirms aerodynamic drag is at least comparable to PR drag for $\mathcal{F} > 1$, thus demonstrating that the conditions necessary to `trigger' a runaway are satisfied rather generically (see also Appendix \ref{app:2}).  Also note that the fact that $t_{\rm run}$ rapidly increases with $x_{\rm thick}$ confirms our implicit assumption that the runaway timescale is dominated by the earliest times when $x_{\rm thick}$ is largest.

For parameters relevant to the above calculation ($b_{\rm PR} = 10^{-2}$; $\mathcal{F} = 2$; $x_{0}$ = 5) equation (\ref{eq:t_run}) predicts a runaway timescale $t_{\rm run} \sim 10^{3}t_{\nu}$, consistent with the results shown in Figure \ref{fig:tauthick} to within a factor $\approx 2$.   

The bottom panel in Figure \ref{fig:mdotthick} shows the fraction of accreted mass as a function of time for the same calculation.  Note that only $\sim 10$ per cent of the total disk mass is accreted during the build-up stage, with the majority instead consumed during the final runaway.  The maximum accretion rate achieved during runaway can be estimated by assuming that the entire solid disk mass $M_{\rm tot}$ (eq.~[\ref{eq:Mtot}]) is consumed on the viscous time $t_{\nu}$ (eq.~[\ref{eq:t_nu}]):
\ba
\dot{M}_{\rm max} \approx \frac{M_{\rm tot}}{t_{\rm visc}} \approx \frac{\tau_{d,0}x_{0}^{2}(\Delta x/x_{0})}{b_{\rm PR}\mathcal{F}}\dot{M}_{\rm PR,in}
\label{eq:mdot_max}
\ea
Unlike in cases without gas feedback, for which the accretion rate is limited to the maximum rate set by PR drag $\dot{M}_{\rm PR,in}$, note that $\dot{M}_{\rm max}$ is proportional to the total disk mass $M_{\rm tot} \propto \tau_{d,0}$ and hence can substantially exceed $\dot{M}_{\rm PR,in}$.  Also note, however, that equation (\ref{eq:mdot_max}) applies only to systems well above the runaway threshold ($\mathcal{F} \gg 1$; see below), for which $t_{\rm run} \ll t_{\rm acc,thick}$ (without gas coupling; eq.~[\ref{eq:t_thick_decoupled}]), such that most of the total mass is indeed accreted during the final runaway.  

\begin{figure}
\resizebox{\hsize}{!}{\includegraphics[angle=0]{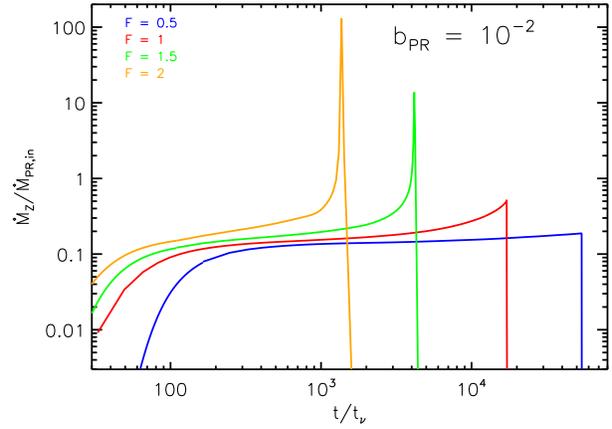}}
\caption[] {Gas accretion rate onto the WD surface $\dot{M}_Z \equiv \dot{M}(x=0)$ (in units of $\dot{M}_{\rm PR,in}$) as a function of time, calculated for several solutions with the same parameters as in Figures \ref{fig:tauthick} and \ref{fig:mdotthick} ($x_{0} = 5$; $b_{\rm PR} = 0.01$; $\tau_{d,0} = 10$) but for different values of the feedback parameter $\mathcal{F} = 0.5$ ({\it blue}), 1 ({\it red}), 1.5 ({\it green}), and 2 ({\it orange}).  Note that the condition $\mathcal{F} \gtrsim 1$ describes well the threshold for a runaway, while the delay time until runaway scales as $t_{\rm run} \propto \mathcal{F}^{-2}$ (eq.~[\ref{eq:t_run}]).
}
\label{fig:mdot_F0.01}
\end{figure}

Figure \ref{fig:mdot_F0.01} shows $\dot{M}_{Z}(t)$, calculated for several models using the same parameters as in Figures \ref{fig:tauthick} and \ref{fig:mdotthick} ($x_{0} = 5$; $b_{\rm PR} = 0.01$; $\tau_{d,0} = 10$) but for different values of the feedback parameter $\mathcal{F}$.  Note that runaway occurs only for $\mathcal{F} \gtrsim 1$, remarkably similar to the threshold predicted by R11b.  Also note that the delay timescale to runaway scales $t_{\rm run} \propto \mathcal{F}^{-2}$, consistent with the analytic prediction in equation (\ref{eq:t_run}).  The rise time of the accretion rate from zero to the PR value during the initial build-up phase also varies with the parameters $b_{\rm PR}$ and $\mathcal{F}$, as determined instead in this case by the optically-thin accretion time $t_{\rm acc,thin} \propto \mathcal{F}^{-1}b_{\rm PR}^{-1}$ (eq.~[\ref{eq:t_thin}]). 

\begin{figure}
\resizebox{\hsize}{!}{\includegraphics[angle=0]{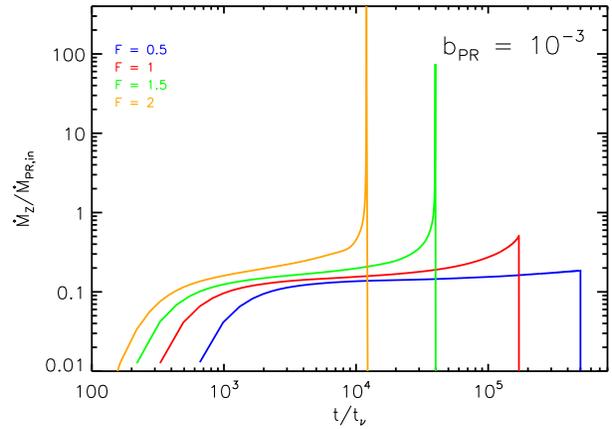}}
\caption[] {Same as Figure \ref{fig:mdot_F0.01}, but calculated for $b_{\rm PR} = 10^{-3}$.  Note that again the threshold condition for runaway is well described as $\mathcal{F} \gtrsim 1$, but that the delay time until runaway $t_{\rm run} \propto b_{\rm PR}^{-1}$ occurs later (at fixed $\mathcal{F}$) due to the lower value of $b_{\rm PR}$ (eq.~[\ref{eq:t_run}]).
}
\label{fig:mdot_F0.001}
\end{figure}

\begin{figure}
\resizebox{\hsize}{!}{\includegraphics[angle=0]{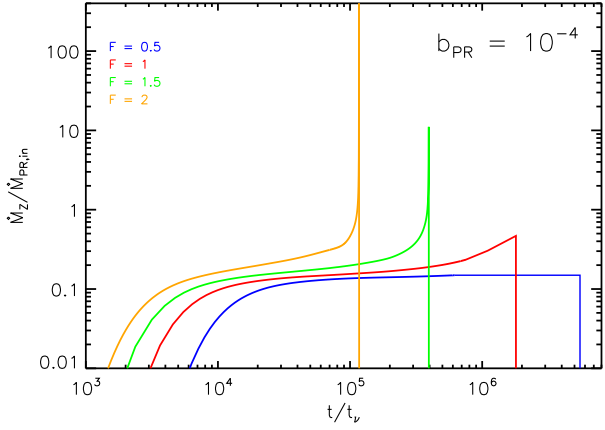}}
\caption[] {Same as Figures \ref{fig:mdot_F0.01} and \ref{fig:mdot_F0.001}, but calculated for $b_{\rm PR} = 10^{-4}$.  }
\label{fig:mdot_F0.0001}
\end{figure}

Figures \ref{fig:mdot_F0.001} and \ref{fig:mdot_F0.0001} show similar results, but now calculated assuming lower values for the PR parameter $b_{\rm PR} = 10^{-3}$ and $b_{\rm PR} = 10^{-4}$, respectively.  In both cases we again find that $\mathcal{F} \gtrsim 1$ is a good estimate of the threshold for runaway, and we confirm that $t_{\rm run} \propto b_{\rm PR}^{-1}\mathcal{F}^{-2}$, as predicted by equation (\ref{eq:t_run}).  

\begin{figure}
\resizebox{\hsize}{!}{\includegraphics[angle=0]{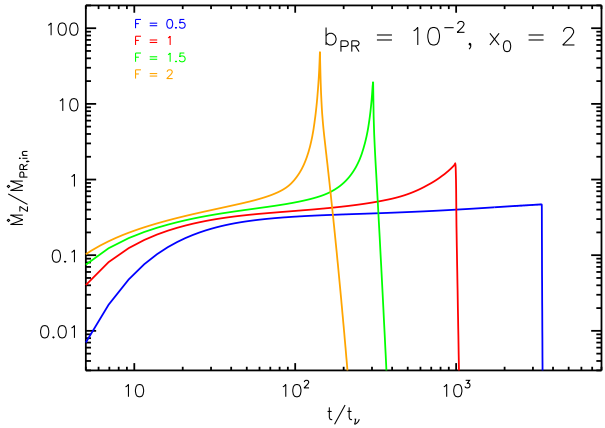}}
\caption[] {Same as Figure \ref{fig:mdot_F0.01} but calculated for an initial gaussian ring with a smaller central radius $x_{0} = 2$. Note that again the runaway threshold occurs at $\mathcal{F} \approx 1$, but that the delay time until runaway $t_{\rm run} \propto x_{0}^{3}$ at fixed $\mathcal{F}$ occurs much earlier (eq.~[\ref{eq:t_run}]). }
\label{fig:mdot_F0.01smallx}
\end{figure}

A final important feature of our results is the sensitive dependence of the runaway timescale on the innermost radius where the disk becomes optically thick $t_{\rm run} \propto x_{\rm thick}^{3}$ (eq.~[\ref{eq:t_run}]).  Figure \ref{fig:mdot_F0.01smallx} shows our results for $\dot{M}_{Z}(t)$ from a series of calculations identical to those shown in Figure \ref{fig:mdot_F0.01}, except calculated for an initial gaussian ring centered at a smaller radius $ x_{0} = 2 \approx x_{\rm thick} $.  The evolution of $\dot{M}_{Z}$ is qualitatively similar to that shown in Figure \ref{fig:mdot_F0.01} for a ring centered at $x_{0} = 5$, except that the runaway time occurs an order of magnitude sooner, consistent with the ratio $\approx (2/5)^{3}$ predicted by equation (\ref{eq:t_run}).

Figure \ref{fig:tauthick}a also shows that when gas drag is important the radial profile of $\Sigma_g$ does not follow a single power law for $r>R_{in}$, as it does without gas drag, see Figures \ref{fig:tauthin} \& \ref{fig:tauthick}b. Instead, $\Sigma_g$ decays slower with $r$ at distances where there is overlap between gas and the optically thick segment of the disk of solids. This is understood to be a direct consequence of the back-reaction of the aerodynamic drag on the gaseous disk, i.e. the non-trivial $\Lambda$ in equation (\ref{eq:gas}).

\subsection{Optically Thick Top-Hat}

\begin{figure}
\resizebox{\hsize}{!}{\includegraphics[angle=0]{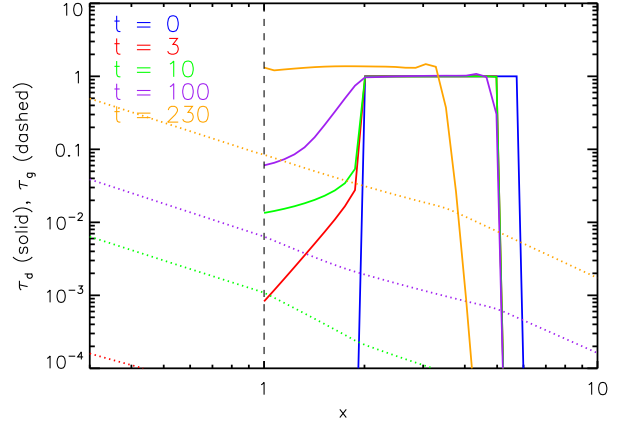}}
\caption[] {Similar to Figures \ref{fig:tauthick}, but now calculated assuming a flat `top-hat' profile between $x = 2$ and $x = 6$ for the initial distribution of optically-thick dust.  The dust ({\it solid lines}) and gas ({\it dotted lines}) density profiles are shown at several times $t = 0$ ({\it blue}), $t = 3$ ({\it red}), $t = 10$ ({\it green}), $t = 100$ ({\it purple}), and $t = 230$ ({\it orange}), in units of the viscous time at $x = 1$.  The calculation is performed for parameters $b_{\rm PR} = 10^{-2}$, $\zeta_{in} = 0.04$, and $\mathcal{F} = 2$.}
\label{fig:tau_tophat}
\end{figure}

\begin{figure}
\resizebox{\hsize}{!}{\includegraphics[angle=0]{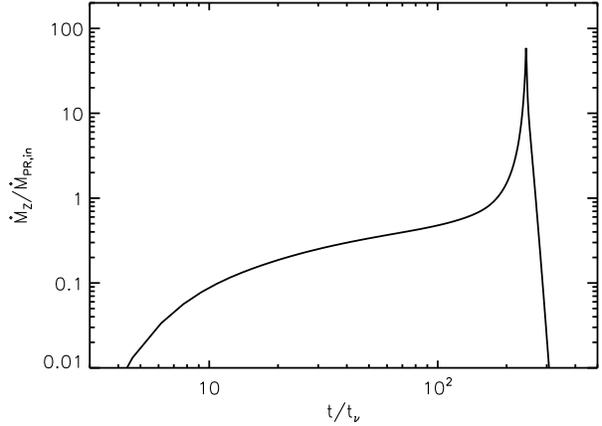}}
\caption[] {Gas accretion rate onto the WD surface $\dot{M}_{Z}$ for the top-hat calculation shown in Figure \ref{fig:tau_tophat}.  Note that the accretion evolution is similar to the case of a gaussian ring with a similar characteristic inner radius (see the $\mathcal{F} = 2$ case in Fig.~\ref{fig:mdot_F0.01smallx}).}
\label{fig:mdot_tophat}
\end{figure}

Our calculations have thus far focused on the idealized case of a gaussian ring for the initial dust radial profile (eq.~[\ref{eq:taud_ring}]).  However, the same qualitative results, including the runaway process, are of general applicability.  To illustrate this point, in Figures \ref{fig:tau_tophat} and \ref{fig:mdot_tophat} we show the gas/dust evolution and gas accretion rate for a calculation which assumes that the initial distribution of solids instead has a `top hat' profile, characterized by a constant value of $\tau_{d} = 1$ from $x = 2$ to $x = 6$.  We again assume characteristic values for the parameters $\mathcal{F} = 2$, $b_{\rm PR} = 10^{-2}$, and $\zeta_{in} = 0.04$.  Though still highly idealized, a top-hat distribution of solids spread over a factor of a few in radii is probably a more accurate description of the initial distribution of solids following the tidal disruption of a minor body.

Figure \ref{fig:tau_tophat} shows that the disk evolves in a qualitatively similar fashion to the gaussian ring shown in Figures \ref{fig:tauthick} and \ref{fig:mdotthick}.  In particular, an optically thin tail develops initially, which later increases due to gas drag, before a runaway occurs at $t = t_{\rm run} \approx 200\,t_{\nu}$ once solids with $\tau_{d} \approx 1$ reach the sublimation radius.  The timescale until runaway can again be reasonably well estimated using the analytic estimate in equation (\ref{eq:t_run}), provided that one replaces the inner radius of the optically-thick ring $x_{\rm thick}$ with the inner radius $x = 2$ of the initial optically-thick top-hat distribution.  

We conclude that the runaway accretion of solids due to gas drag is an inevitable fate for optically-thick disks in the presence of strong feedback $\mathcal{F} \gtrsim 1$.  With everything else fixed, the timescale for runaway depends on the minimum radius of the optically-thick material $x_{\rm thick}$ (eq.~[\ref{eq:t_run}]), while the maximum accretion rate depends on the mass of the solid disk (eq.~[\ref{eq:mdot_max}]).

\section{Applications and Observational Tests}
\label{sect:applications}

In this section we present applications of our calculations and observational implications of the runaway accretion model.


\subsection{True Vs. `Measured' Accretion Rate}
\label{sect:settle}

\begin{figure}
\resizebox{\hsize}{!}{\includegraphics[angle=0]{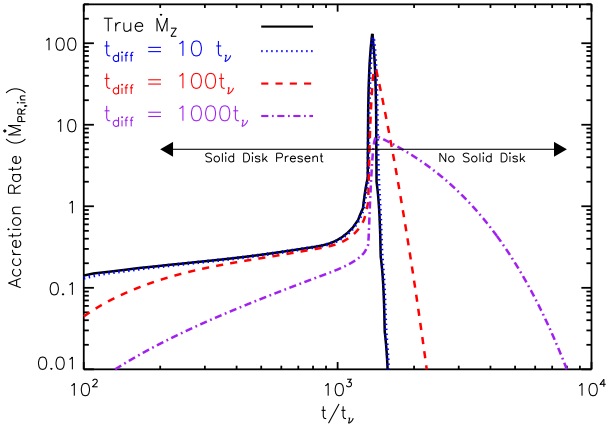}}
\caption[] {Actual gas accretion rate $\dot{M}_{Z}$ from our fiducial solution shown in Figures \ref{fig:tauthick}a and \ref{fig:mdotthick} ({\it solid line}) compared to the `measured' accretion rate $\dot{M}_{\rm obs,Z}$ calculated using the steady-state approximation (eq.~[\ref{eq:mdot_measured}]), shown for different assumptions about the value of $t_{\rm diff}$ = 10 $t_{\nu}$ ({\it blue dotted line}), 10$^{2}$ $t_{\nu}$ ({\it red dashed line}), and 10$^{3}$ $t_{\nu}$ ({\it green dot-dashed line}), where $t_{\nu}$ is the accretion timescale at the sublimation radius.  See text for details.
}
\label{fig:mdot_compare}
\end{figure}

Because the timescale required for heavy elements to diffuse below the outer convective zone of the WD is much shorter than the WD age, metals observed at the surface indicate recent pollution by an external source, such as accretion from a gaseous disk.  Under the joint action of accretion and gravitational settling, the total mass of metals in the convective zone $M_{Z_{i}}$ with charge $Z_{i}$ evolves according to (e.g.~\citealt{Koester&Wilken06})
\begin{eqnarray}
\frac{dM_{Z_{i}}}{dt} = \dot{M}_{Z_{i}} - \frac{M_{Z_{i}}}{t_{\rm set,i}},
\label{eq:Mdot_conv}
\end{eqnarray}
where $\dot{M}_{Z_{i}} = \dot{M}_{Z}X_{Z_{i}}$, $X_{Z_{i}}$ is the mass fraction of metal $Z_{i}$ in the accreting gas, and $t_{\rm set,i}$ is the settling (or diffusion) time of element $Z_{i}$, which depends on the temperature and composition of the WD (e.g.~\citealt{Paquette+86}).  A typical WD of mass $M_{\rm WD} = 0.6M_{\odot}$, for instance, requires $\sim 2\times 10^{8}$ yrs and $6\times 10^{8}$ yrs to cool to a temperature $T \simeq 1.5\times 10^{4}$ K and $T \simeq 1.0\times 10^{4}$ K, respectively \citep{Bergeron+95}.  In a H-rich atmosphere (DAZ WDs), over this time interval the settling times of most observed elements increase from $t_{\rm set,i}\sim$ days to $\sim 10^{3}$ years.  For a He-rich atmosphere (DBZ WD), by contrast, the convective zone is much larger, resulting in a much longer settling time $t_{\rm set,i} \approx 10^{5.5-6}$ yrs, which depends less sensitively on temperature \citep{Paquette+86}.

Gas is fed from the sublimation radius to the WD surface on the viscous time $t_{\nu}$ (eq.~[\ref{eq:t_nu}]).  If $t_{\rm set,i} \ll t_{\nu}$, then the atmospheric composition reaches a steady state $dM_{Z_{i}}/dt = 0$, in which the mass in the convective zone is approximately given by $M_{Z_{i}} \approx \dot{M}_{Z_{i}}t_{\rm set,i}$.  Since the surface composition $M_{Z_{i}}$ is measured via spectral line diagnostics and theoretical models of the total mass in the WD convective zone, and $t_{\rm set,i}$ is also calculable from the known properties of the WD, then the total instantaneous gas accretion rate in this limit is well estimated by the expression (e.g.~\citealt{Koester&Wilken06}; \citealt{Farihi+09})
\begin{eqnarray}
\dot{M}_{Z} \simeq  \frac{M_{Z_{i}}}{X_{Z_{i}}t_{\rm set,i}} \equiv \dot{M}_{\rm Z,obs} \quad (t_{\rm set,i} \ll t_{\nu}). 
\label{eq:mdot_measured}
\end{eqnarray}
Equation (\ref{eq:mdot_measured}) provides a reliable estimate of $\dot{M}_{Z}$ in the case of hot DAZ WDs (which easily satisfy $t_{\rm set,i} \ll t_{\nu}$), but it can be grossly inaccurate in the opposite limit that $t_{\rm set,i} \gtrsim t_{\nu}$, as may apply to DBZ or cool DAZ WDs.  

Figure \ref{fig:mdot_compare} illustrates this point in the context of our models by showing a comparison of the actual gas accretion rate $\dot{M}_{Z}$ from our fiducial solution from Figure \ref{fig:tauthick} to the `measured' value $\dot{M}_{\rm Z,obs}$ one obtains by naively applying equation (\ref{eq:mdot_measured}), the latter shown for different assumptions about the ratio $t_{\rm set,i}/t_{\nu} = 10,10^{2},10^{3}$.  We calculate $M_{Z}(t)$ by integrating equation (\ref{eq:Mdot_conv}) directly, assuming that $X_{Z,i} = 1$ and using $\dot{M}_{Z,i}$ derived from our calculations shown in Figure \ref{fig:mdotthick}.

In the case that $t_{\rm set,i} \lesssim 10t_{\nu}$, Figure \ref{fig:mdot_compare} shows that `measured' accretion rate $\dot{M}_{\rm Z,obs}$ indeed accurately reflects the true instantaneous accretion rate.  On the other hand, when $t_{\rm set,i} \gg 10t_{\nu}$, $\dot{M}_{Z}$ and $\dot{M}_{\rm Z,obs}$ may differ substantially.  In particular, prior to the time when $\dot{M}_{Z}$ peaks, $\dot{M}_{\rm Z, obs}$ {\it underestimates} the true accretion rate by a factor up to $\sim 10$.  Since this epoch corresponds to when the solid disk is still present, our results suggest that one should exercise caution when interpreting the accretion rates in those systems with an observed IR excess when $t_{\rm set,i} \gg t_{\nu}$.  On the other hand, at times after $\dot{M}_{Z}$ peaks, $\dot{M}_{\rm Z,obs}$ instead {\it overestimates} the true accretion rate.  This discrepancy is one explanation for those systems shown in Figure \ref{fig:Farihi} with high measured accretion rates $\dot{M}_{\rm Z,obs}$, yet no IR excess indicating the current presence of a solid disk.   

One concrete application of the results in Figure \ref{fig:mdot_compare} is to cool DBZ WDs with large $t_{\rm set,i}$, for which one predicts that $\dot{M}_{\rm Z,obs}$ underestimates the true current accretion rate $\dot{M}_Z$ for systems with detected near-IR excesses (indicating a current solid disk), whereas $\dot{M}_{\rm Z,obs}$ instead overestimates $\dot{M}_Z$ in DBZ WD systems without solid disks.  Note also that since the {\it ratio} of the inferred accretion rates of different elements is subject to similar uncertainties, one should also exercise caution when using similar methods to infer the precise composition of the accreted bodies in systems for which $t_{\rm set,i} \gtrsim t_{\nu}$ (cf.~\citealt{Xu&Jura11}).  Previous works have attempted to generalize estimates of $\dot{M}_{Z}$ to cases when $t_{\rm set,i} > t_{\nu}$ (or to constrain the entire accretion history of the event), by adopting simplified assumptions about the time evolution of the accretion rate (e.g.~\citealt{Jura+09a}), e.g.~$\dot{M} \propto \exp[-t/t_{\rm acc}$], where $t_{\rm acc}$ is the accretion timescale.  However, given the complicated accretion evolution in cases when feedback is effective, we again urge caution when adopting such simplified models.    

We conclude with one final point regarding what inferences can be drawn about the past accretion history of a WD using the currently `measured' accretion rate $\dot M_{\rm Z_i,obs}\equiv X_{\rm Z_i}\dot M_{\rm Z,obs}$ given by equation (\ref{eq:mdot_measured}).  Equation (\ref{eq:Mdot_conv}) can be trivially integrated to give the mass of metals with charge $Z_i$ in the WD atmosphere as a function of time, provided that the evolution of the true accretion rate $\dot M_{\rm Z_i}(t)$ is known:
\ba
M_{\rm Z_i}(t)=\int\limits_0^t\dot M_{\rm Z_i}(t^\prime)
e^{-(t-t^\prime)/t_{\rm set,i}}dt^\prime,
\label{eq:acc_history}
\ea
where we have assumed that $M_{Z_i}=0$ at time $t=0$. If we define $\dot M_{\rm Z_i,max}(<t)$ to be the maximum value reached by the true accretion rate $\dot M_{\rm Z_i}$ in the past, then one estimates that
\ba
M_{\rm Z_i}(t)&<&\dot M_{\rm Z_i,max}(<t)\int\limits_0^t
e^{-(t-t^\prime)/t_{\rm set,i}}dt^\prime
\nonumber\\
&=&
\dot M_{\rm Z_i,max}(<t)t_{\rm set,i}\left(1-e^{-t/t_{\rm set,i}}\right),
\label{eq:acc_history_limit}
\ea
such that for any $t$ one has
\ba
\dot M_{\rm Z_i,obs}(t)=\frac{M_{\rm Z_i}(t)}{t_{\rm set,i}}<\dot M_{\rm Z_i,max}(<t).
\label{eq:mdor_constraint}
\ea
Thus, the currently `measured' value of $\dot M_{\rm Z_i,obs}$ always provides a {\it lower limit} on the maximum $\dot M_{\rm Z_i}$ achieved in the past.  Conversely, for any given `measured' value of $\dot M_{\rm Z_i,obs}$ one can deduce that during some epoch in the past the {\it true} mass accretion rate $\dot M_{\rm Z_i}$ was {\it at least as high} as $\dot M_{\rm Z_i,obs}$.


\begin{figure}
\resizebox{\hsize}{!}{\includegraphics[angle=0]{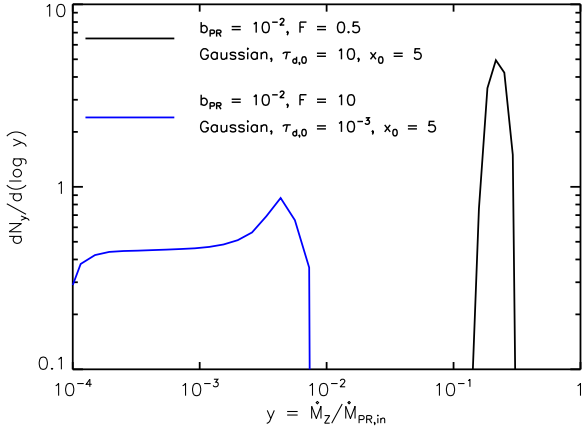}}
\caption[] {Probability distribution $dN_{y}/d($log $y)$ for detecting a WD debris disk when the accretion rate is $y = \dot{M}_{Z}/\dot{M}_{\rm PR,in}$, calculated for two solutions {\it without} runaway accretion.  A blue line shows the distribution calculated for the optically thin disk with $\tau_{d,0} = 10^{-3}$ from Figures \ref{fig:tauthin} and \ref{fig:mdotthin}.  A black line shows the distribution calculated for the optically-thick disk ($\tau_{d,0} = 10$) with a feedback parameter $\mathcal{F} = 0.5$ too low for runaway accretion.  In both cases the accretion rate peaks below the characteristic value set by PR drag, which is insufficient to explain the highest measured WD accretion rates shown in Figure \ref{fig:Farihi}. 
}
\label{fig:mdot_histo5}
\end{figure}

\begin{figure}
\resizebox{\hsize}{!}{\includegraphics[angle=0]{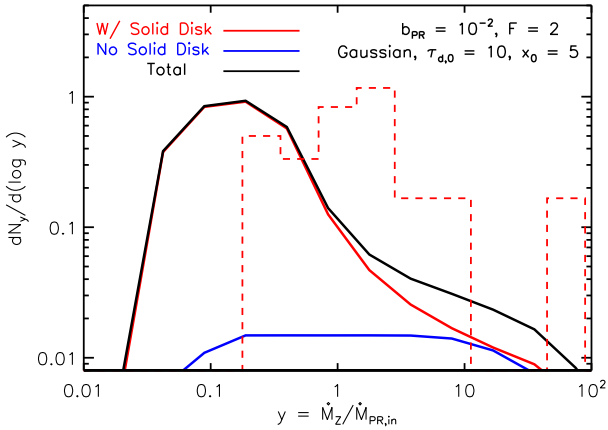}}
\caption[] {Probability distribution $dN_{y}/d($log $y)$ for detecting a WD debris disk when the accretion rate is $y = \dot{M}_{Z}/\dot{M}_{\rm PR,in}$, calculated for the evolution of the optically-thick gaussian ring ($\tau_{d,0} = 10$; $x_{0} = 5$; $\Delta x = 0.5$) with runaway accretion $(\mathcal{F} = 2)$ shown in Figures \ref{fig:tauthick} and \ref{fig:mdotthick}.  Note that the runaway process results in a wider distribution, extending to much higher accretion rates than in the otherwise identical calculation without runaway accretion ($\mathcal{F} = 0.5$) shown in Fig.~\ref{fig:mdot_histo5}.    
}
\label{fig:mdot_histo1}
\end{figure}

\begin{figure}
\resizebox{\hsize}{!}{\includegraphics[angle=0]{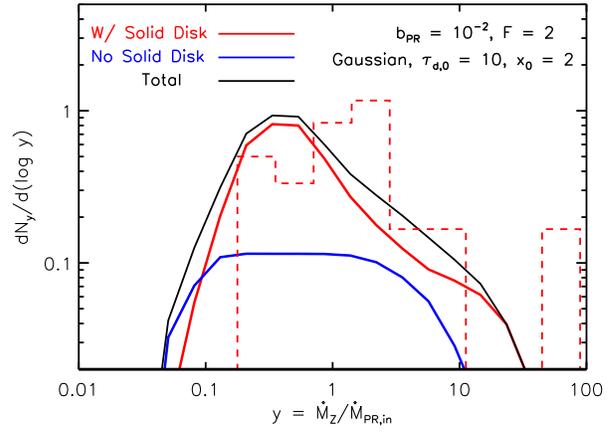}}
\caption[] {Same as Figure \ref{fig:mdot_histo1}, but calculated using the gaussian ring centered at the smaller central radius $x_{0} = 2$ (Fig.~\ref{fig:mdot_F0.01smallx}).
}
\label{fig:mdot_histo4}
\end{figure}

\begin{figure}
\resizebox{\hsize}{!}{\includegraphics[angle=0]{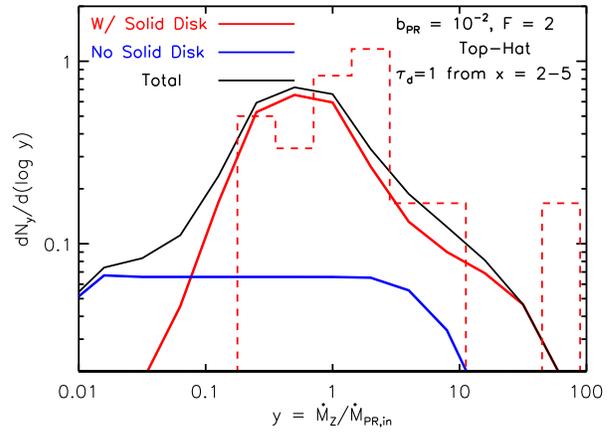}}
\caption[] {Same as Figure \ref{fig:mdot_histo1} but calculated using the solution for the `top hat' initially density profile (Figs.~\ref{fig:tau_tophat} and \ref{fig:mdot_tophat}).
}
\label{fig:mdot_histo3}
\end{figure}


\subsection{Accretion Rate Distribution}
\label{sect:distribution}

Given a complete model for the evolution of WD debris disks, one can begin to address observational questions, such as the probability of detecting a WD debris disk system when the accretion rate has some particular value.  Figures \ref{fig:mdot_histo5}-\ref{fig:mdot_histo4} show several such examples of the normalized distribution $dN_{y}/d($log $ y)$ of observed accretion rates $y = \dot{M}_{Z}/\dot{M}_{\rm PR,in}$.  Each distribution is created using a single disk evolution calculation from $\S\ref{sect:results}$, under the highly idealized assumptions that (1) all WD debris disks are identical to the model from which each distribution was created; and (2) the system is equally likely to be detected at any moment during active gas accretion.  Although neither of these assumptions is likely valid, these distributions nevertheless inform what kinds of debris disk systems could contribute to the observed distribution of accretion rates shown in Figure \ref{fig:Farihi}.

Figure \ref{fig:mdot_histo5} shows the distribution calculated for two cases {\it without} runaway accretion: the optically-thin ($\tau_{d,0} = 10^{-3}$) gaussian ring shown in Figures \ref{fig:tauthin} and \ref{fig:mdotthin} ({\it blue line}); and the optically-thick ($\tau_{d,0} = 10$) gaussian ring without strong feedback ($\mathcal{F} = 0.5$) shown in Figures \ref{fig:tauthick} and \ref{fig:mdotthick} ({\it black line}).  In both cases the distribution is fairly narrowly concentrated about the characteristic accretion rate set by PR drag acting alone.  Accretion rates $\dot{M}_{Z} \lesssim \dot{M}_{PR, in}$ at the low end of the measured values in Figure \ref{fig:Farihi} are thus consistent with resulting from a superposition of such systems, as may result e.g.~from a range in the masses or orbital parameters of disrupted asteroids.  However, disk evolution without runaway accretion cannot explain the highest observed accretion rates $\dot{M}_{Z} \sim 10-100\dot{M}_{\rm PR,in}$.   

Figures \ref{fig:mdot_histo1}-\ref{fig:mdot_histo3} show the accretion rate distribution ({\it solid black line}) for several disk evolutions {\it with} runaway accretion.  In these cases we further divide the distribution between those times when the solid disk is still present ({\it solid red line}) and those times after it has completely sublimated ({\it solid blue line}).  Also plotted for comparison with a dashed red line is the normalized distribution of measured WD metal accretion rates shown in Figure \ref{fig:Farihi} from the sample of \citet{Farihi+09,Farihi+10b}, including just those systems with an IR excess.  Note that our results correspond to the true accretion rate $\dot M_Z$, while the observational sample uses the `measured' rates $\dot M_{\rm Z,obs}$ (see \S \ref{sect:settle}). Thus, one should exercise caution when comparing them.

Figure \ref{fig:mdot_histo1} shows the fiducial case of an optically-thick gaussian disk ($\tau_{d,0} = 10$, $x_{0} = 5$; $\mathcal{F} = 2$; $b_{PR} = 10^{-2}$), as shown in Figures ~\ref{fig:tauthick}a and \ref{fig:mdotthick}.  The probability distribution in this case still peaks at a relatively low accretion rate $\dot{M}_{Z} \sim 0.1-1\dot{M}_{\rm PR,in}$ set by the characteristic range in accretion rate during the `build-up' to runaway.  However, the final stages in runaway process produces a substantial `tail' in the distribution extending to much higher accretion rates $\dot{M}_{Z} \sim 10-100 \dot{M}_{\rm PR,in}$. 

Although small number statistics and potential selection effects introduce many uncertainties, note the overall qualitative similarity between the shapes of the theoretical and observed accretion rate distributions.  Somewhat better agreement with the observed distribution is obtained for disks formed with more compact inner radii, as shown in Figures \ref{fig:mdot_histo4} and \ref{fig:mdot_histo3} for the cases of a gaussian ring centered at $x_{0} = 2$ (Fig.~\ref{fig:mdotthick}) and the top hat distribution with $\tau_{d} = 1$ from $x = 2$ to $x = 5$ (Fig.~\ref{fig:mdot_tophat}), respectively.  The top hat distribution in particular illustrates the broad accretion rate distribution produced by even an ensemble of identical systems in the (most physically realistic) case of a radially-extended disk that undergoes runaway accretion.
 
Figures \ref{fig:mdot_histo1}-\ref{fig:mdot_histo3} also illustrate that the probability of detecting actively accreting systems when the solid disk is still present ({\it solid red line}) is much greater than after the solid disk has entirely sublimated ({\it solid blue line}).  We caution, however, against attaching much significance to a direct comparison between our predictions for the fraction of actively accreting systems without solid disks and the fraction of metal-rich WDs with no detected IR excess, given potential selection effects and other ways that IR emission could be difficult to detect even in systems with disks (see Fig.~\ref{fig:mdot_histo5} and $\S\ref{sect:discussion}$).  




\section{Discussion}
\label{sect:discussion}


Our results demonstrate that runaway accretion is a ubiquitous feature in the evolution of WD debris disks with strong gas-solid coupling and/or weak gas viscosity.  We find that the condition $\mathcal{F} \equiv t_{\nu}/t_{s} \gtrsim 1$ for runaway predicted by R11b using local calculations is a surprisingly accurate diagnostic of the runaway threshold. 

In Appendix \ref{app:2} we present and analyze a simple global analytical model of the coupled evolution of the two disks which makes two simplifying assumptions: (a) that the back-reaction of aerodynamic drag on the gaseous disk can be neglected and (b) that the disk of solids is always optically thick at $R_{in}$. Even though this model is fully global and far more realistic compared to the calculations done in R11b we are still able to derive from it analytically the runaway condition (eq.~[\ref{eq:runaway_condition}]), which is essentially analogous to ${\cal F}\gtrsim 1$. This additionally reaffirms the robustness of the runaway threshold found in R11b and this work.

Is $\mathcal{F} > 1$ physically realizable in WD debris disk systems?  If one adopts our standard model for the aerodynamic drag between solid and gaseous disks described in $\S\ref{subsect:aero}$, then satisfying this condition $\mathcal{F} \gtrsim 1$ requires gas viscosity $\alpha_{in} \lesssim 10^{-2}-10^{-4}$ near the inner edge of the disk (see eq.~[\ref{eq:Fnum}] and surrounding discussion).  At first glance this condition appears unlikely to be satisfied because the gas temperature is sufficiently high $T_{g} \gg 10^{3} K$ that thermal ionization should be efficient, whereas observations of fully-ionized accretion disks \citep{King+07} and numerical simulations  of the magneto-rotational instability (MRI) in ideal MHD (e.g.~\citealt{Fromang&Papaloizou07}; \citealt{Davis+10}) typically suggest higher values of $\alpha \sim 0.01-0.1$.  

Note, however, that gas and solids overlap at the radii $\gtrsim R_{s}$ in our model and {\it Spitzer} observations reveal that micron-size dust grains are present in many systems \citep{Jura+09b}.  The presence of small dust in the disk can significantly lower the ionization fraction below the nominal equilibrium value because small grains are very efficient at absorbing charge.  Even though the observationally inferred mass of micron-size Si grains ($\lesssim 10^{18}$ g, \citep{Jura+09b}) is much less than that of a sizeable asteroid thought to give rise to a solid disk, small grains may still dominate the total surface area of particles.  A lower free charge density increases the resistivity of the gas, which in turn reduces the effective turbulent viscosity by suppressing the MRI due to non-ideal MHD effects (e.g.~\citealt{Fleming+00}).  The relevant value of $\alpha_{in}$ in our model could thus be much lower than would be expected in a fully ionized disk, thus making strong feedback $\mathcal{F} \gg 1$ much more likely.  

In addition to lowering the viscosity, the runaway condition $\mathcal{F} \gg 1$ could also be satisfied if the interaction between gas and solids is stronger than is assumed in our baseline model in $\S\ref{subsect:aero}$.  Below we describe two such additional mechanisms for solid-gas coupling: enhanced aerodynamic drag due to an eccentric gaseous disk ($\S\ref{subsect:ecc}$) and `inductive coupling' by a magnetic field threading the disk ($\S\ref{sect:Bfield}$).  Accounting for these additional possibilities, it appears quite possible that the conditions for runaway accretion are actually achieved in Nature.

Our model provides useful predictions for the properties of gaseous disks fed by sublimation which may be verified by observations.  In particular, equation (\ref{eq:sigma_g}) suggests a characteristic estimate of the gas surface density at $R_{in}$ for $\dot M_Z\sim \dot M_{\rm PR}$.  The value of $\Sigma_g$ at this location can easily be higher than $10^{-2}$ g cm$^{-2}$ if $\dot M_Z$ exceeds $\dot{M}_{\rm PR}$, as should be the case during runaway (but note that it is the true rate $\dot M_Z$ that sets $\Sigma_g$, see \S\ref{sect:settle} for details).  We also point out that viscous evolution in our model naturally results in the profile of $\Sigma_g(r)$ which is shallower for $r<R_{in}$ than for $r>R_{in}$, see equation (\ref{eq:Sig2}) and Figure \ref{fig:schematic}.  This implies that the gas mass inside the cavity of the disk of solids is lower than one would naively assume, which might help explain the observed lack of line emission from this region of the gaseous disk \citep{Melis+10}.

The lifetime of disks that undergo runaway accretion is set predominantly by the delay timescale until runaway $t_{\rm run}$ given in equation (\ref{eq:t_run}).  Substituting the values of $t_{\nu}$ (eq.~[\ref{eq:t_nu}]) and $b_{\rm PR}$ (eq.~[\ref{eq:PR_par}]) into this expression one finds that
\ba
t_{\rm life} \sim t_{\rm run} \simeq 5\times 10^{5}\mathcal{F}^{-1}\left(\frac{x_{\rm thick}}{2}\right)^{3}\left(\frac{a}{\rm cm}\right){\,\rm yr} ,
\label{eq:t_life} \qquad (\mathcal{F} > 1)
\label{eq:tlife2}
\ea
where we have adopted characteristic values for the gas temperature $T_{g} = 5\times 10^{3}$ K, solid particle density $\rho_{d} = 2.5$ g cm$^{-3}$, and mean molecular weight $\mu = 28m_{p}$.  We normalize the inner edge of the optically thick disk to a characteristic value $x_{\rm thick} \approx 2$ which best reproduces the observed WD accretion rate distribution (Fig.~\ref{fig:mdot_histo4}).  Note that $t_{\rm life}$ is independent\footnote{As shown in Figure \ref{fig:drag_picture} our drag law underestimates aerodynamic coupling in the optically thin case. In reality coupling is stronger and this may lower the disk lifetime to some extent.} of the (uncertain) strength of the disk viscosity $\alpha_{in}$ and the precise form of the drag force between the solids and gas (except implicitly through the value of $\mathcal{F}$).  

The lifetimes of WD debris disk are poorly constrained by observations, but one can set a rough upper limit of $t_{\rm life} \lesssim 10^{5}-10^{6}$ yr (\citealt{Kilic+08}; \citealt{Farihi+09};  Farihi, private communication).  Equation (\ref{eq:t_life}) shows that this constraint translates into an upper limit on the characteristic particle size $a \lesssim (0.2-2)\mathcal{F}$ cm.  For $\mathcal{F} > 1$ (as required for runaway to occur in the first place) one thus requires particles with a typical size $a \lesssim$ several cm, consistent with (relatively weak) existing observational constraints (e.g.~\citealt{Graham+90}; \citealt{Jura+07b}).

An important outstanding question is why some WDs with high inferred accretion rates have no detected IR excess.  Our calculations in $\S\ref{sect:distribution}$ predict that the total fraction of actively accreting systems without solid disks should be relatively small, typically $\lesssim$ 20 per cent.  As discussed in $\S\ref{sect:settle}$, one possible explanation in the case of systems with long metal diffusion times $t_{\rm set,i} \gg t_{\nu}$ (as characterize WDs with cool temperatures and/or He-enriched atmospheres) is the discrepancy between the true instantaneous WD accretion rate and the value inferred using atmospheric abundances and the steady-state approximation.  Figure \ref{fig:mdot_compare} shows that for $t_{\rm set,i} \gtrsim 10^{2}-10^{3}t_{\nu}$, the accretion rate inferred using the steady-state model can remain at a high level for a timescale comparable to or longer than the disk lifetime itself even after the solid disk has completely sublimated.  

Another possible explanation for high-$\dot{M}_{Z}$ systems without an IR excess is that a solid disk is in fact present but cannot be detected, for instance because of a poor viewing angle and/or because the disk is concentrated in a narrow ring of width $\Delta r \approx 0.01-1R_{\odot}$ (\citealt{Farihi+10b}).  Our calculations in Figures \ref{fig:tauthick} and \ref{fig:tau_tophat} confirm that this is a viable possibility by showing that optically-thick disks largely maintain their original width, until the end of the disk lifetime when the runaway process is nearly complete. 

We conclude by addressing several extensions to our standard model and outstanding issues to be explored further in future work.


\subsection{Effects of Non-Circular Gas Motion}
\label{subsect:ecc}

Our calculation of aerodynamic drag in \S\ref{subsect:aero} explicitly assumed that both the solid and gaseous material move on purely {\it circular} orbits, in which case the difference in azimuthal velocity between the disks $v_{\phi,rel}$ is due solely to the pressure support in the gaseous disk (eq.~[\ref{eq:v_rel}]).  Since this velocity difference is relatively small (compared to, say, the orbital velocity), then the resultant aerodynamic coupling between the disks is not very strong.  For this reason runaway accretion may require special circumstances, such as a low viscosity in the gaseous disk.

At the same time, however, the Ca II emission lines from some WD systems are observed to have asymmetric shapes (\citealt{Gansicke+06,Gansicke+07}).  These have been interpreted as due to {\it non-circular} motions in the gaseous disk, with \citet{Gansicke+06} for instance fitting the Ca II line profiles from WD SDSS 1228+1040 as arising from a disk with eccentricity $e\approx 0.02$. If this interpretation is correct, then the assumption of purely circular gas motion is not justified in at least some WD systems.  

If the gas disk is eccentric (even if the origin of the eccentricity is unknown) then one must re-evaluate the drag force calculation from $\S\ref{subsect:aero}$ to account for non-circular motions.  In Appendix \ref{app:ecc} we carry out such a calculation, in particular demonstrating that in the small-$e$ limit ($e\to 0$) the orbit-averaged azimuthal force can be written by analogy with equation (\ref{eq:aerodrag2}) as $\langle f_{a}(r)\rangle_\varphi=A_e\Sigma_g(r)$, see equation (\ref{eq:force_av_small_e}), where now the coupling constant is
\ba
A_e=\frac{e^2}{8\mbox{Re}_\star}\frac{r^2\Omega^3}{c_s}.
\label{eq:A_e}
\ea 
The ratio of $A_e$ to the coupling constant $A_a$ defined by 
equation (\ref{eq:A_a}) is
\ba
\frac{A_e}{A_a}=\frac{e^2}{8}\left(\frac{\Omega r}{c_s}\right)^4.
\label{eq:A_rat}
\ea
As a result, even a small eccentricity of the gaseous streamlines 
$e\gtrsim (c_s/\Omega r)^2$ (which is about $10^{-6}$ at
the the inner disk edge) is sufficient for non-circular gas motions to dominate over the azimuthal velocity difference resulting due to gas pressure support for purposes of coupling the gaseous and solid disks. 

Using equations (\ref{eq:feedback}) and (\ref{eq:A_e}) we can estimate the feedback parameter for coupling due to the non-circular gas motions:
\ba
&& {\cal F}_e=\frac{4A_{e,in} R_{in}}{\alpha_{in} c_s^2}
=\frac{e^2}{2\alpha_{in}\mbox{Re}_\star}
\left(\frac{\Omega r}{c_s}\right)^3
\label{eq:feedback2}\\
&& \approx 30\left(\frac{\alpha_{in}}{10^{-3}}\frac{{\mbox{Re}_\star}}{10^{2}}\right)^{-1}\left(\frac{e}{10^{-4}}\right)^2
\left(\frac{r}{0.2R_{\odot}}\right)^{-3/2},
\nonumber
\ea
where we have again adopted fiducial values for $\mu = 28m_{p}$, $T_{g} = 5000$ K and $M_{\star} = 0.6M_{\odot}$.

This expression shows that non-circular gas motions are extremely efficient at driving the runaway evolution of the debris disks, with even slightly eccentric ($e\sim 10^{-4}$) gas motions resulting in very strong feedback.  In fact, this coupling may be {\it too} effective in the sense that ${\cal F}\gg 1$ implies a very short timescale $\sim t_s={\cal F}^{-1}t_\nu$ on which the particulate disk evolves due to coupling during the runaway. The duration of the build-up phase $t_{\rm run}$ is also lowered for ${\cal F}\gg 1$, see equation (\ref{eq:tlife2}). Measurements of line asymmetries in WD gaseous disks typically imply $e\gg 10^{-4}$ if one interprets asymmetry as due to eccentric gas motions (e.g.~$e\approx 0.02$ for SDSS 1228+1040 [\citealt{Gansicke+06}] and $e\approx 0.2-0.4$ for SDSS 0845+0855 [\citealt{Gansicke+07}]).  In the latter case the inferred eccentricity would be so high as to result in relative velocity between the gas and particle disks $\sim$ hundreds of km s$^{-1}$.  How such a highly supersonic velocity difference could be maintained in practice is not clear, unless the orbits of the solid particles are extremely closely aligned with the eccentric trajectory of the gas, which appears unlikely because of dissipative collisions between particles (see \S \ref{subsect:gas_in_coll}).  It is furthermore unclear how the particle disk could avoid being destroyed by the enormous aerodynamic drag in this system, or by sputtering of the particle surfaces by the gas. 

These considerations lead us to believe that the line asymmetries observed in the spectra of gaseous WD disks result not from actual non-circular gas motions, but rather as the result of a non-axisymmetric surface brightness over the face of the disk.  Non-axisymmetric brightness patterns, resulting e.g. due to spiral waves or other instabilities which produce an azimuthally varying surface density distribution or line excitation conditions, could explain the observed line asymmetries without invoking highly non-circular gas motions.  \citet{Hartmann+11} demonstrate that this is indeed possible. 

To summarize, we conclude that the actual deviations of gas motions from purely circular in WD disks are probably not as dramatic as has been inferred based on eccentric disk models, and most likely correspond to $e\lesssim 10^{-4}$.  Equation (\ref{eq:feedback2}) nevertheless shows that even weakly non-circular gas motions could easily result in drag sufficient to produce runaway disk evolution, thus supporting the general picture outlined in \citet{Rafikov11b}.  Additional work is clearly required to identify what mechanisms are capable of driving such non-circular gas motions.


\subsection{White Dwarf Magnetic Field}
\label{sect:Bfield}

We have thus far neglected the possible effects of the WD magnetic field on the disk evolution.  If the WD field is sufficiently strong, it may, for instance, disrupt the gaseous disk above the WD surface, channeling matter onto the surface along the magnetic poles, as in magnetic cataclismic variables.  

Assuming that at large radii the WD magnetic field is dipolar $B(r) = B_{\star}(r/R_{\star})^{-3}$ with a surface field strength $B_{\star}$, then the magnetosphere and disk stresses balance at the Alfven radius $R_{A}$ given by (e.g.~\citealt{Ghosh&Lamb78})
\ba
R_{A} &\simeq& \left(\frac{3B_{\star}^{2}R_{\star}^{6}}{2\dot{M}\sqrt{GM_{\star}}}\right)^{2/7} \nonumber \\ &\simeq& \nonumber 1.2R_{\odot}\left(\frac{B_{\star}}{\rm kG}\right)^{4/7}\left(\frac{\dot{M}}{10^{8}\rm\,g\,s^{-1}}\right)^{-2/7}\left(\frac{R_{\star}}{10^{-2}R_{\odot}}\right)^{12/7}\left(\frac{M_{\star}}{0.6M_{\odot}}\right)^{-1/7}
\label{eq:Ra}
\ea
Equation (\ref{eq:Ra}) shows that for typical accretion rates in WD debris disks $\dot{M}_{Z} \sim 10^{6}-10^{10}$ g s$^{-1}$, fields as small as $\sim 0.1-1$ kG are sufficient to affect the flow near the sublimation radius $R_{\rm in} \sim 0.2R_{\odot}$.  

The surface magnetic field strengths of isolated WDs inferred from Zeeman (spectro-)polarimetry show a wide range of values, from `high field' WDs with $\gtrsim 10^{4}$ G \citep{Wickramasinghe&Ferrario00}, to only upper limits of $\lesssim 1-100$ kG in most other systems (e.g.~\citealt{Schmidt&Smith95}; \citealt{Gansicke+07}).  Overall the distribution of field strengths appears to be well characterized by a power-law distribution $dN/dB \propto B^{\alpha}$ with index $\alpha \simeq -1.5$ for $B > 100$ kG (\citealt{Fabrika&Valyavin99}).  However, the peak and average of the distribution both occur at lower fields than can be measured and hence are not well constrained.  Although few DZ WDs have measured magnetic fields, note that at least a few systems, G165-7 and G77-50, have strong surface fields $B_{\star} \sim 650$ kG (\citealt{Dufour+06}) and $\sim 120$ kG (\citealt{Farihi+11b}), respectively.

It thus seems possible that the WD magnetic fields could affect a significant fraction of debris disks.\footnote{Caveats include that (1) current measurements constrain the total surface field, of which the dipole may only be a minor component; (2) the magnetic field could be systematically lower in actively accreting systems if, for instance, the field is `buried' by diamagnetic screening (e.g.~\citealt{Cumming+01}).}  If a dynamically important magnetic field threads the gaseous or solid disk, this may complicate the picture of disk evolution presented in this paper.  If $R_{\star} \lesssim R_{A} \lesssim R_{s}$ then matter interior to the Alfven radius is channeled directly onto the WD surface, creating an inner `hole' in the gaseous disk.  If $R_{A} \gtrsim R_{s}$, then matter may be placed onto field lines as soon as it sublimates at $r \approx R_{s}$.  Since viscous spreading is no longer required to carry away the angular momentum of the accreted gas (which is instead extracted by magnetic torques and used to spin-up the WD), then the picture of overlapping gas and solid disks, and the resulting model of runaway accretion developed in this paper, might no longer remain valid.  
An important caveat, however, is if small dust particles in the disk suppress the ionization fraction of the gas, as discussed earlier in this section.  In this case the magnetic field may not couple efficiently to the gaseous disk, negating the influence of even a strong field.  A more detailed model of the coupled evolution of small dust, gas-field coupling, and the MRI in the sublimation layer will be required to address the flow structure in this region and the resulting influence on the accretion evolution.  

If the WD magnetic field is sufficiently strong to disrupt the gaseous disk, then it may also affect the rate of solid accretion via `inductive' coupling (\citealt{Drell1965,Gurevich1978}). Because the WD rotates at a different (typically much lower) angular velocity than the accretion disk, this differential rotation produces an electric field in frame of the rotating particles of order ${\bf E \simeq -(v_{\phi}}/c){\bf \times B}$, where ${\bf v}_{\phi}$ is the velocity at which the magnetic field sweeps through particles.  If the WD magnetosphere is loaded with dense enough plasma supplying abundant free charges then this electric field not only polarizes debris particles but also induces a DC current through them. This current in turn couples to the magnetic field, resulting in the azimuthal drag force on the solid particles. The strength of this inductive coupling depends on the conductivity of particles and is rather uncertain. We leave a more detailed analysis of this coupling mechanism to future work.


\subsection{Problem of Condensation}
\label{sect:condensation}

The spatial coincidence of gaseous and particulate debris disks around WDs poses an interesting problem.\footnote{We are grateful to Bruce Draine for directing our attention to this issue.}  Quite generally one expects that the debris particles and gas will have the same chemical composition, since the latter originates from the former (by evaporation in our model, or by sputtering as in \citealt{Jura08}).  It is then not clear how two separate phases of the same material --- solid and gaseous --- can exist simultaneously in stable phase equilibrium.

Indeed, the temperature of the solid particles is necessarily below the sublimation temperature $T_{s}$, while the surrounding gas is likely to be hotter than $T_s$ \citep{Melis+10}.  Atoms of gas should stick upon colliding with the surface of debris particles, resulting in condensation with a probability equal to the accommodation coefficient $\alpha_{acc}<1$.  The rate of condensation from the gas phase per unit area of the solid surface is $\sim \alpha_{acc}\rho_g c_s\sim \alpha_{acc}\Omega\Sigma_g$ (assuming a solid debris disk with $\tau_d\sim 1$), such that the gas disk should be depleted on a timescale $\sim \Omega^{-1}\alpha_{acc}^{-1}$.  Even if $\alpha_{acc}$ is as small as $\sim 10^{-2}$ (its value is typically higher; see \citealt{Leitch-Devlin&Williams85}), then the entire gaseous disk should still condense within several hundred orbital time scales.  The latent heat released during this process may somewhat slow down the rate of condensation, but ultimately cannot prevent it.

Observations showing that the gaseous and solid components overlap in WD debris disks (\citealt{Brinkworth+09}; \citealt{Melis+10}) demonstrate that this simple logic is somehow flawed.  However, at the moment we do not have a satisfactory explanation for this puzzle of condensation.

\subsection{Production of Metallic Gas by Particle Collisions}
\label{subsect:gas_in_coll}

Our work assumes that all metallic gas in the disk results from the sublimation of solids at the inner edge of the particulate disk (in $\S$\ref{sect:disk_solid} we allowed for the possibility of gas condensation at larger radii as described in $\S\ref{sect:condensation}$, but we did not explore this possibility in detail).  However, previous authors (\citealt{Jura08}; \citealt{Farihi+09}) have suggested that gas can also be produced by high-velocity collisions between solid particles within the debris disk.  These authors argue that, due to the large orbital speed $\sim$ hundreds km s$^{-1}$, even moderately non-circular particle motions (corresponding to orbital eccentricities as low as $\sim 10^{-2}$) result in particle collisions with sufficiently high velocity of several km s$^{-1}$ to effectively sputter, or even vaporize, solids into metallic gas.  This process of gas creation was envisioned to operate continuously throughout the entire solid disk, thereby naturally explaining the observed radial coincidence\footnote{In our model this overlap is naturally achieved as a result of viscous spreading of the gaseous disk (see \S\ref{sect:model}).} between solid and gaseous disks (\citealt{Melis+10}).

We believe that this mechanism is unlikely to be an important source of gas production, even on timescales short compared to the disk lifetime.  Observations indicate (\citealt{Jura03}; \citealt{Jura+07a}) that WD debris disks are typically optically thick, i.e. $\tau_d\gtrsim 1$.  It is well known (\citealt{GT78}; \citealt{Farihi+08b}) that under such circumstances the characteristic timescale between the pairwise particle collisions is comparable to the local orbital period, which is less than an hour in the present context.  If particles are indeed continuously being converted to gas via binary collisions, then the whole disk of solid debris should be converted into gas within just a handful of collision timescales, i.e. $\sim$ day!  This is in contradiction with the fact that the NIR excesses observed around many WDs are persistent for at least decade-long timescales.

Even if gas production via collisions is highly inefficient, such that only a small fraction of the solid mass is converted into gas in every collision (as would be necessary to extend the lifetime of the solid disk), then collisions are still incapable of providing a steady source of gas.  This is because high relative velocities between particles are quickly damped (again, on a timescale of several orbits) due to the inelastic nature of particle collisions \citep{GT78}.  As a result, the solid disk rapidly settles into a dynamically cold configuration, similar to the rings of Saturn (for which the particle velocity dispersion is only $\sim 0.01-0.1$ cm s$^{-1}$; \citealt{Cuzzi+10}), thereby eliminating the possibility of additional destructive {\it high-velocity} particle collisions.

Despite these objections, collisional gas production is likely to be important during and immediately following the tidal disruption of the asteroid-like body by the WD, which could last for several orbital timescales of the incoming body, i.e. $\sim$ tens to hundreds of years.  During this stage, collisions may occurs between large solid objects at relative speeds of many km s$^{-1}$, resulting in a large fraction of the total solid mass being converted into gas.  Some of this gas will re-condense onto the surfaces of surviving particles, with the rest accreting onto the WD on the [relatively short] viscous timescale (see eq.~[\ref{eq:t_nu}]).

To summarize, the arguments presented above strongly argue against the possibility that {\it sustained, long-term} (comparable to the disk lifetime $\sim 10^5-10^6$ yr) gas production in WD debris disks results from inelastic solid particle collisions.


\section{Summary}
\label{sect:summary}

\begin{table}
\begin{center}
\vspace{0.05 in}\caption{Summary of the key properties of different regimes of accretion in WD debris disks.$^{\dagger}$}
\label{table:models}
\resizebox{9cm}{!}
{
\begin{tabular}{lcc}
\hline
\hline
\multicolumn{1}{c}{Accretion Regime} &
\multicolumn{1}{c}{Disk Lifetime} &
\multicolumn{1}{c}{Peak $\dot{M}_{Z}$} 
\\
& ($t_{\nu}\times b_{PR}^{-1}\mathcal{F}^{-1}$)$^{\ddagger}$ & ($\dot{M}_{\rm PR,in}$) \\
\hline 
\hline
Optically Thin ($\tau_{d} \ll \zeta$) & $x_{0}^{2}\times\zeta_{in}$ & $\zeta_{in}^{-1}\tau_{d}$ \\ 
\\[1pt]
Optically Thick ($\tau_{d} \gg 1$) & & \\
\,\,\,\,\,\,Weak Feedback ($\mathcal{F} < 1$) & $x_{0}^{2}\Delta x \times \tau_{d} $  & $x_{0}^{-1}$ \\ 
\,\,\,\,\,\,Strong Feedback ($\mathcal{F} > 1$) & $x_{0}^{3}\times \mathcal{F}^{-1}$  & $x_{0}\Delta x\times \tau_{d} b_{\rm PR}^{-1}\mathcal{F}^{-1}$ \\
\hline
\hline
\end{tabular}
}
\end{center}
{$^{\dagger}$Expressions are given in terms of (1) characteristic radius $x_{0}$ and radial thickness $\Delta x \lesssim x_{0}$ of the initial solid disk, in units of the inner edge of the disk $R_{\rm in} \approx R_{\rm s} \sim 0.2R_{\odot}$ (eq.~[\ref{eq:R_S}]); (2) viscous time $t_{\nu}$ at the inner disk edge (typically $\sim 10^{3}$ yr, depending on viscosity $\alpha_{in}$; eq.~[\ref{eq:t_nu}]); (3) characteristic accretion rate $\dot{M}_{\rm PR,in}$ due to optically-thick PR drag (typically $\sim 10^{8}$ g s$^{-1}$; eq.~[\ref{eq:dotM_PR_phys}]); (4) dimensionless parameters introduced in $\S\ref{sect:dim-less}$, including the feedback parameter $\mathcal{F}$ (eqs.~[\ref{eq:feedback},\ref{eq:Fnum}]), PR-parameters $b_{PR}$ (typically $\sim 10^{-4}-1$; eqs.~[\ref{eq:def2},\ref{eq:bPR}]), and the incidence angle of WD radiation $\zeta_{in}$ (typically $\sim 0.02$; eq.~[\ref{eq:def2}]).  $^{\ddagger}$ We express time in the useful combination $t_{\nu}b_{PR}^{-1}\mathcal{F}^{-1}$, which takes on the numerical value $\sim 6\times 10^{4}(a/{\rm cm})$ yr for typical parameters (independent of the strength of the viscosity or the mechanism coupling solids and gas), where $a$ is the characteristic particle size.}
\end{table}

The growing number of `debris disks' detected around WDs provides a unique opportunity to study the physics of the interaction between solid and gas disks in a way complementary to the study of normal debris disks around pre-main sequence stars.  As a likely repository for minor bodies originating from the outer stellar system, these disks also provide an indirect probe of the dynamics of planetary systems around evolved stars.  As such, they also indirectly inform our knowledge of the fraction of the intermediate-mass stars harboring planets.   

One of the biggest mysteries regarding known WD debris disk systems is the wide range of metal accretion rates, in particular those extending to much higher values than can be explained by Poynting-Robertson drag alone (see Fig.~\ref{fig:Farihi}).  In this paper we have begun to address this and other issues using global calculations of the coupled evolution of gaseous and solid components of WD debris disks.  Table \ref{table:models} summarizes the key properties of different regimes of accretion in WD debris disks.  Our primary conclusions are summarized as follows:
\begin{itemize}

\item{The observed coexistence of gas and dust in WD debris disks (\citealt{Melis+10}) naturally results from viscous spreading of gas created at the sublimation radius.}
\item{Drag between the gaseous and solid disks leads to an enhanced accretion rate over that set by PR drag alone.  Our results in $\S\ref{sect:results}$ show that the disk evolution in the presence of strong aerodynamic drag is dramatically altered from cases without drag (see Fig.~\ref{fig:tauthick}-\ref{fig:mdot_tophat}).  We confirm the mode of `runaway' accretion predicted by R11b when the rate of drag-induced accretion exceeds the rate at which viscosity spreads gas away.  The threshold for runaway is well described by the condition $\mathcal{F} \equiv t_{\nu}/t_{s} \gtrsim 1$ (eq.~[\ref{eq:feedback}]) for a variety of different configurations of solids (see also Appendix \ref{app:2}).}
\item{The runaway process is characterized by two stages (Fig.~\ref{fig:mdotthick}): (1) an initial `build-up' phase, during which an optically-thick disturbance with $\tau_{d} \sim 1$ moves inwards from the inner edge of the optically thick disk; (2) a `run-away' phase characterized by super-exponential increase in the accretion rate, once $\tau_{d} \sim 1$ is satisfied near the sublimation radius.}  
\item{The predicted distribution of measured accretion rates given an ensemble of optically-thick WD debris disks undergoing evolution leading to runaway is broadly consistent with the shape of the current observational sample (Figs.~\ref{fig:mdot_histo1}-\ref{fig:mdot_histo3}).  Most observed systems are predicted to have accretion rates in the range $\dot{M}_{Z} \sim (0.1-1)\dot{M}_{\rm PR,in}$, but a smaller tail of the distribution comprising $\sim 10-20$ percent of systems should extend to higher accretion rates.}
\item{Satisfying the runaway condition $\mathcal{F} > 1$ in WD debris disks requires a low value for the disk viscosity $\alpha_{in} \lesssim 10^{-3}-10^{-4}$ and/or strong solid-gas coupling (eq.~[\ref{eq:Fnum}]).  A low viscosity could result if the MRI is suppressed due to low ionization caused by the presence of small dust grains.}
\item{The runaway condition $\mathcal{F} > 1$ could also be satisfied even for higher viscosity if the strength of solid-gas coupling is stronger than our baseline model for aerodynamic drag presented in $\S\ref{subsect:aero}$.  In $\S\ref{subsect:ecc}$ and Appendix $\ref{app:ecc}$ we show that even a mildly eccentric gaseous disk $e \gtrsim 10^{-4}$ results in a stronger drag force than our baseline model.  These results also suggest that the observed asymmetries in WD emission lines are unlikely to result from highly eccentric disks, but instead may result from non-axisymmetric disturbances in the disk surface brightness (e.g.~spiral wave patterns).}
\item{The disk lifetime in the runaway accretion model is set by the delay timescale during the ``build-up'' phase $t_{\rm life} \approx t_{\rm run}$ (eq.~[\ref{eq:t_run}]).  Reconciling this lifetime with the observational upper limits on the disk lifetime $t_{\rm life} < 10^{5}-10^{6}$ yrs places an upper limit $a \lesssim$ several cm on the characteristic particle size (eq.~[\ref{eq:tlife2}]).}
\item{The steady-state assumption commonly adopted to calculate WD metal accretion rates (eq.~[\ref{eq:mdot_measured}]) may be inaccurate when the settling time of metals in the WD convective zone $t_{\rm set,i} \gg 10t_{\nu}$, resulting in some cases in an over- or under-estimate of the true instantaneous accretion rate by an order of magnitude or more (Fig.~\ref{fig:mdot_compare}).  This is a possible explanation for even heavily metal-rich WDs with no current IR excess in cases for which the metal diffusion time is comparable to the disk lifetime.}

\end{itemize}

\section*{Acknowledgments}
BDM is supported by NASA through Einstein Postdoctoral Fellowship grant number PF9-00065 awarded by the Chandra X-ray Center, which is operated by the Smithsonian Astrophysical Observatory for NASA under contract NAS8-03060.  Financial support for RRR is provided by the Sloan Foundation, NASA via grant NNX08AH87G, and NSF via grant AST-0908269.


\bibliographystyle{mn2e}
\bibliography{ms}

\begin{appendix}


\section{Mass Accretion Due to Collisional Viscosity}
\label{app:collisional}

Here we evaluate the role of `internal' sources of viscosity within the particulate disk, similar to those thought to characterize the evolution of the rings of Saturn, on the rate of solid accretion in WD debris disks.  Angular momentum transport in a dense ring is caused by three effects (\citealt{Wisdom&Tremaine88}; \citealt{Takeda&Ida01}): (1) translational transport due to the particle motions; (2) collisional transport; and (3) transport due to the gravitational torques, which are important only when self-gravity is important. 

According to \citet{Daisaka+01}, the efficiency of gravitational transport is determined by two parameters: (1) the optical depth of the particulate disk $\tau_d$ (eq.~[\ref{eq:sig0}]); and (2) the ratio of the mutual Hill radius $r_h\equiv r(2m_p/3M_\star)^{1/3}$ to twice the particle radius $a$: 
\ba
r_h^* &\equiv &\frac{r_h}{2a}=\left(\frac{\pi}{9}\frac{\rho}
{M_\star}\right)^{1/3}r 
\nonumber\\
&\approx & 0.1\left(\frac{\rho_{d}}{\mbox{2.5 g cm}^{-3}}
\frac{0.6M_\odot}{M_\star}\right)^{1/3}\left(\frac{r}{0.2~R_\odot}\right),
\label{eq:rhstar}
\ea
where $m_p = (4\pi/3)\rho_{d}a^{3}$, $\rho_{d}$, and $r$ are the particle mass, density, and distance from the WD, respectively.  Note that $r_h^*$ is independent of the particle mass. 

Equation (\ref{eq:rhstar}) shows that $r_h^*\lesssim 0.5$ is typically satisfied for the range of radii $r \lesssim R_\odot$ relevant to WD debris disks.  In this regime gravitational transport can be neglected provided that (\citealt{Daisaka+01})
\ba
\tau_d \lesssim  \tau_{sg}\approx \frac{0.08}{\left(r_h^*\right)^3} \approx 
60\left(\frac{M_\star}{0.6 M_\odot}\frac{2.5~\mbox{g cm}^{-3}}{\rho_{d}}\right)
\left(\frac{0.2~R_\odot}{r}\right)^3.
\label{eq:self-grav}
\ea

If gravitational transport is negligible then internal transport is instead dominated by the collisional viscosity $\nu\approx a^2\Omega \tau_d$ (\citealt{Daisaka+01}), where $\Omega = (GM_{\star}/r^{3})^{1/2}$.  The resulting mass flux can be estimated as 
\ba
\dot M_{Z}&=&3\pi\nu\Sigma_d = \pi^{-1}m_{p}\tau_{d}^{2}\Omega \nonumber \\
&\approx& 10~\mbox{g s}^{-1}
\left(\frac{m_p}{1~\mbox{g}}\right)\left(\frac{\tau_{d}}{60}\right)^{2}\left(\frac{M_{\star}}{0.6M_{\odot}}\right)^{1/2}\left(\frac{r}{0.2R_{\odot}}\right)^{-3/2}.
\label{eq:dotM_nsg}
\ea

Equation (\ref{eq:dotM_nsg}) shows that in order to explain typical values $\dot M_Z\sim 10^8$ g s$^{-1}$ of the accretion rates onto metal-rich WDs, then for\footnote{For $\tau_d\lesssim 1$, as characterize e.g. Saturn's rings, one would need $\gtrsim 10$ m particles to achieve the same value of $\dot M_Z$.} $\tau_d \sim 60$ one requires a typical particle mass $m_{p} \sim 10^7$ g, corresponding to an object of size $a \gtrsim 1$ m.  Obviously, particles of size $a \sim 1$ m in a disk with optical depth $\tau_{d} \sim 30$ cannot reside in a monolayer configuration; rather, the vertical thickness of the disk must be $\sim 100$ m, thus requiring a total solid mass $\sim (4\pi/3)\rho_{d}a r^2\tau_d \approx 10^{25}$ g (assuming characteristic disk size $r = 0.2$ R$_\odot$).  This mass is similar to that of a single 2000 km basaltic asteroid, which is significantly higher than the total mass of high-Z elements typically inferred to pollute metal-rich WDs (\citealt{Farihi+10a}).

Explaining the highest inferred values of $\dot M_Z\sim 10^{10}-10^{11}$ g s$^{-1}$ requires even larger disk particles ($\sim 10$ m) and, hence, even more massive disks.  In principle the ring viscosity is enhanced if $\tau_d$ is sufficiently high that equation (\ref{eq:self-grav}) is violated; when self-gravity becomes important, the viscosity instead becomes $\nu\propto \tau_d^3$ (\citealt{Daisaka+01}).  However, the required disk mass in this case approaches that of a terrestrial size body, an unlikely explanation for all WDs with high $\dot M_Z$.   We conclude that viscosity due to internal stresses within the particulate ring is unlikely to play an important role in the majority of metal-rich WDs.  A similar conclusion was reached by \citet{Farihi+08b}, who also estimated accretion times based on an analogy with planetary rings, finding that typically they greatly exceed the WD cooling timescale.  



\section{Evolution of accretion disks with mass sources/sinks: 
analytical solutions.}
\label{app:1}

Here we analyze equation (\ref{eq:gas}) to derive some general 
results for the viscous evolution of 
an accretion disk with sources or sinks of mass and power law 
dependence of viscosity on radius
\ba
\nu=\nu_0 r^n,
\label{eq:PL_nu}
\ea
but without sources or sinks of the angular momentum, i.e. 
$\Lambda=0$ in equation (\ref{eq:gas}). The initial condition 
for the disk evolution is
\ba
\Sigma_g(r,t=0)=\Sigma_{g,0}(r).
\label{eq:init_cond}
\ea

Power law viscosity is 
realized in the conventional model of $\alpha$-viscosity 
(\citealt{Shakura&Sunyaev73}) provided that the gas temperature $T_{g}$
has a power law dependence on radius: 
\ba
\nu(r)=\frac{\alpha c_s^2}{\Omega}=
\alpha\frac{k_B T(r)r^{3/2}}{\mu\sqrt{GM_\star}}.
\label{eq:alpha}
\ea
In particular, if $T_{g}(r)$=const then $n=3/2$. If $T(r)\propto 
r^{-1/2}$, as expected for optically thin gas in equilibrium with stellar
heating, then $n=1$.

Following \citet{Tanaka11} and introducing new time-like and spatial 
coordinates $T$ (not to be confused with temperature) and $w$
\ba
T\equiv 3\nu_0 t,~~~~~w(r)\equiv \frac{2}{2-n}r^{1-n/2},
\label{eq:new_coords}
\ea
and looking for solution in the form 
\ba
\Sigma_g(r,t)=r^{-n-1/4}\sigma_g(w,T),
\label{eq:Sigma_anzatz}
\ea
equation (\ref{eq:gas}) can be transformed to 
\ba
\frac{\partial\sigma_g}{\partial T} &=&
\frac{\partial^2\sigma_g}{\partial w^2}
+\frac{1}{w}\frac{\partial\sigma_g}{\partial w}
-\frac{l^2}{w^2}\sigma_g+\Psi(w,T),~~~~
l=\frac{1}{2(2-n)},
\label{eq:Bessel}
\nonumber\\
\Psi(w,T)&\equiv & [r(w)]^{n+1/4}\frac{S_g(r(w),T)}{3\nu_0}.
\label{eq:Psi}
\ea

We now introduce Hankel transforms $\tilde F=\left\{\tilde \sigma_g(k,T), 
\tilde \Psi(k,T), \tilde \Sigma_{g,0}(k)\right\}$ for each of the 
functions $\tilde F=\left\{\sigma_g,\Psi,\Sigma_{g,0}\right\}$ according to 
formula
\ba
\tilde F(k,T)&=&\int_0^\infty F(w,T)J_l
(kw)w dw,\nonumber\\
F(w,\tau)&=&\int_0^\infty \tilde F(k,T)J_l(kw)kdk,
\label{eq:Hankels}
\ea
Plugging these relations into (\ref{eq:Bessel}) one arrives at the
following simple differential equation for $\tilde\sigma_g(k,T)$:
\ba
\frac{\partial\tilde\sigma_g(k,T)}{\partial T}+
k^2\tilde\sigma_g(k,T)=\tilde\Psi(k,T)
\label{eq:dif}
\ea
with initial condition
\ba
\tilde\sigma_g(k,\tau=0)=\tilde\Sigma_{g,0}(k).
\label{eq:sig_init}
\ea

Solving this equation and performing inverse Hankel transform on
$\tilde\sigma_g$ we obtain
\ba
\Sigma_g(r,t)&=& r^{-n-1/4}\int\limits_0^\infty\left[\tilde\Sigma_{g,0}(k)+
\int\limits_0^T\tilde\Psi(k,T^\prime)e^{k^2T^\prime}
dT^\prime\right]
\nonumber\\
&\times & e^{-k^2T}J_l(kw(r))kdk.
\label{eq:soln}
\ea
We can further simplify this solution by plugging in the expressions
for $\tilde\Sigma_{g,0}$ and $\tilde\Psi$ from (\ref{eq:Hankels}), rearranging
the order of integration and using integral 6.633(2) from \citet{Gradshteyn+07}. This gives us
\ba
\Sigma_g(r,t)&=&\int\limits_0^\infty\left(\frac{r(w^\prime)}{r}\right)^{n+1/4}
\Bigg[\Sigma_{g,0}(w^\prime)G(w(r),w^\prime,T)
\nonumber\\
&+&
\left.
\int\limits_0^T\frac{S_g(w^\prime,T^\prime)}{3\nu_0}
G(w(r),w^\prime,T-T^\prime)dT^\prime\right]w^\prime dw^\prime,
\label{eq:soln_short}
\ea
where
\ba
G(w,w^\prime,z)\equiv\frac{1}{2z}\exp\left(-\frac{w^2+w^{\prime 2}}{4z}
\right)I_l\left(\frac{w w^\prime}{2z}\right),
\label{eq:Green}
\ea
and $T(t)$ and $w(r)$ are given by equation (\ref{eq:new_coords}).


\subsection{Singular mass source.}

For our current setup (sublimation at a single radius) it is useful to 
consider the situation when the injection of mass happens over a small 
radial span of the disk at $r=R_{in}$, so that one can 
write
\ba
S_g(r,t)=S_t(t)\delta(r-R_{in}),~~~~~~~S_t(t)=\frac{\dot M(r=R_{in},t)}
{2\pi R_{in}},
\label{eq:mdot_sing}
\ea
see equation (\ref{eq:gas_source}).
For simplicity we will assume that initially disk does not contain 
any mass, i.e. $\Sigma_{g,0}=0$. Abandoning this assumption would amount 
to including the first term in equation (\ref{eq:soln_short}) into 
the consideration.

Plugging equation (\ref{eq:mdot_sing}) into equation (\ref{eq:soln_short})
and introducing $w_0\equiv w(R_{in})$ one finds
\ba
\Sigma_g(r,t)=\frac{2}{2-n}\frac{R_{in}^{5/4}r^{-n-1/4}}{3\nu_0}
\int\limits_0^T S_t(T^\prime)G(w(r),w_0,T-T^\prime)
dT^\prime.
\label{eq:Sig_sing}
\ea

One can obtain a simple solution for $\Sigma_g(r,t)$ at late times 
$T\gg w_0^2$ under the assumption that $S_t(T)$ varies 
slowly. In this case we can take $S_t(T)$ to be approximately 
constant and take it out of the integral in equation 
(\ref{eq:Sig_sing}). Then one gets
\ba
\Sigma_g(r,t) &=& \frac{2S_t}{2-n}\frac{R_{in}^{5/4}r^{-n-1/4}}{3\nu_0}
\int\limits_0^T G(w(r),w_0,z)dz\nonumber\\
&\approx & 
\frac{S_tR_{in}^{5/4}r^{-n-1/4}}{3(2-n)\nu_0}
\int\limits_0^\infty \frac{dz}{z}\exp\left(-\frac{w^2+w_0^2}{4z}
\right)I_l\left(\frac{w w_0}{2z}\right).
\label{eq:Sig_sing_simple}
\ea
Taking the last integral with the aid of relation
\ba
\int\limits_0^\infty \frac{du}{u}e^{-\lambda u}I_\nu(u)=
\frac{1}{\nu}\left(\lambda+\sqrt{\lambda^2-1}\right)^{-\nu}
\ea
one finds that
\ba
\Sigma_g(r,t) &=& \frac{2S_t}{3\nu_0}\frac{R_{in}^{3/2}}{r^{n+1/2}},
~~~r>R_{in},\nonumber\\
&=& \frac{2S_t}{3\nu_0}\frac{R_{in}}{r^{n}},
~~~r<R_{in}.
\label{eq:Sig2}
\ea

With this expression for the surface density one can easily determine
mass accretion rate through the disk:
\ba
\dot M(r)=6\pi r^{1/2}\frac{\partial}{\partial r}\left(\nu\Sigma 
r^{1/2}\right)=0,~~~r>R_{in},
\label{eq:Mdotform}
\ea
while for $r<R_{in}$ one finds $\dot M(r)=2\pi S_t R_{in}$. Thus, all of the
mass injected at $R_{in}$ goes towards the central mass at late times.

\section{Coupled evolution in the absence of momentum feedback 
on the gaseous disk}
\label{app:2}

The goal of this Appendix is to gain analytical insight into the coupled evolution of gaseous and solid WD disks, which are linked by mass exchange due to the evaporation of solids, as expressed by the source term in equation (\ref{eq:gas_source}). 

\begin{figure}
\resizebox{\hsize}{!}{\includegraphics[angle=0]{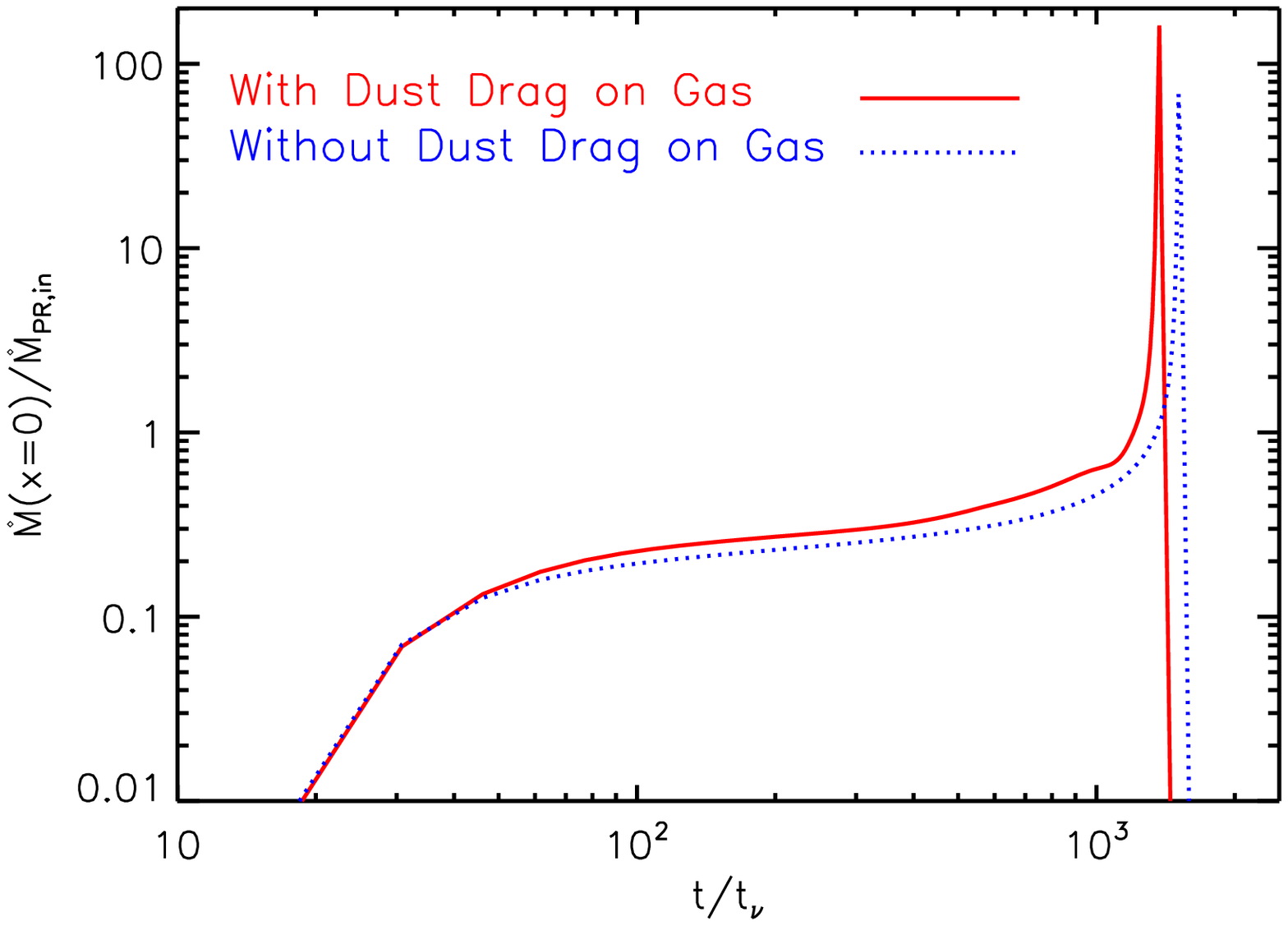}}
\caption[] {Comparison of the evolution of the gas accretion rate $\dot{M}_{Z}(t)$ in two cases for otherwise fully identical simulations: when the back-reaction of the aerodynamic drag on the gas disk is not included ({\it dotted line}) and when it is fully accounted for ({\it solid line}). Note that only small quantitative differences distinguish the behavior of $\dot{M}_{Z}(t)$ in the two cases.
}
\label{fig:nogascouple}
\end{figure}

Here we account for the effect of gas drag on the evolution of the surface density of solids (as we do in $\S\ref{sect:results}$), {\it but we neglect the back-reaction of the drag force on the gas disk evolution}, assuming that its effect is small compared to that of the viscous stresses.  With this simplification the evolution of the gas disk can be studied using the results of Appendix \ref{app:1}, provided that the viscosity obeys the power law ansatz (eq.~[\ref{eq:PL_nu}]).  Figure \ref{fig:nogascouple} shows a comparison between the mass accretion rate $\dot M_{Z}$ computed with and without the back-reaction of the aerodynamic drag on the gaseous disk.  The similar qualitative evolution in both cases justifies our assumption that the back-reaction force on the gas disk can be neglected to first order.

Another important assumption we make here is that the surface density of solids at $R_{in}$ is sufficiently high that $\tau_d(R_{in},t)\gtrsim 1$ is satisfied at all times. This approximation is likely to be valid only during the late stages of the disk evolution, as illustrated in Figures \ref{fig:tauthick}a and 
\ref{fig:tau_tophat}. Thus, the model we present below does not capture the early, ``build-up'' phase of the gaseous disk evolution, but it should work well at late times, during the runaway phase.  This important simplification makes it possible to treat the evolution of the gas disk analytically, separate from the calculation of the solid disk evolution. 

Using equations (\ref{eq:dotM})-(\ref{eq:dotM_a}), (\ref{eq:t_s}) and (\ref{eq:mdot_sing}) one obtains the following expression for the intensity of the mass source feeding the gaseous disk:
\ba
S_t(t)=\frac{\dot M_{PR,in}}{2\pi R_{in}}
+\frac{\eta|\eta|}{2}\frac{R_{in}}{t_s}\Sigma_{g}(R_{in},t),
\label{eq:S_t}
\ea
where we have assumed that $r=R_{in}$ and that $\tau_d(R_{in},t)\gtrsim 1$. 

This expression, coupled with the solution in equation (\ref{eq:Sig_sing}), provides a closed set of equations for the evolution of $\Sigma_g(r,t)$ which can be computed numerically.  Note that a knowledge of $\Sigma_d(r,t)$ is not required to calculate $\Sigma_g(r,t)$ as long as $\tau_d(R_{in},t)\gtrsim 1$. This is because the mass flux of solid debris at $R_{in}$ saturates at the value independent of $\tau_d(R_{in},t)$ when the disk of solids is optically thick.

To demonstrate the conditions under which runaway behavior becomes possible we concentrate on evaluating the time evolution of $\Sigma_g$ at $R_{in}$ only.  By substituting equation (\ref{eq:S_t}) into equation (\ref{eq:Sig_sing}), setting $r=R_{in}$, and casting all variables into dimensionless form with the aid of equations (\ref{eq:def2})-(\ref{eq:PR_par}), one obtains the following integral equation for the evolution of the gas optical depth at the inner edge of the disk $\tau_{g,in}(t)\equiv\tau_g(R_{in},t)$:
\ba
\tau_{g,in}(\tilde t)={\cal F}\frac{2-n}{4}\int\limits_0^{\tilde t}
\left[\eta|\eta|\tau_{g,in}(\tilde t^\prime)+b_{PR}\right]
\tilde G\left(\frac{3(2-n)^2}{4}(\tilde t-\tilde t^\prime)
\right)d\tilde t^\prime,
\label{eq:int_eqn}
\ea
where 
\ba
\tilde G(z)\equiv\frac{1}{2z}\exp\left(-\frac{1}{2z}
\right)I_l\left(\frac{1}{2z}\right).
\label{eq:Gtilde}
\ea
Note that $t_\nu=R_{in}^{2-n}/\nu_0$ for the power law viscosity (\ref{eq:PL_nu}).

Our numerical results presented in \S \ref{sect:results} demonstrate that there are two possibilities for the gas disk evolution: either (1) $\Sigma_g$ saturates at essentially a constant level on timescales long compared to $t_\nu$ or (2) it grows with time in a runaway fashion.  In both cases the behavior of $\Sigma_g$ is determined primarily by the most recent history of the disk and is largely insensitive to conditions in the distant past.  This allows us to extend the lower limit of integration in equation (\ref{eq:int_eqn}) to $-\infty$ for $\tilde t\gg 1$.  This results (upon an obvious change of variables) in the following equation describing the late time behavior of $\Sigma_g$:
\ba
\tau_{g,in}(\tilde t)={\cal F}\frac{2-n}{4}\int\limits_0^{\infty}
\left[\eta|\eta|\tau_{g,in}(\tilde t-z)+b_{PR}\right]
\tilde G\left(\frac{3(2-n)^2}{4}z
\right)dz.
\label{eq:late}
\ea

In the case of orderly behavior, $\tau_{g,in}$ saturates at the constant level, such that the expression in brackets in equation (\ref{eq:late}) can be taken out of the integral.  Using the fact that $\int_0^\infty\tilde G(z)dz=1/(2l)=(2-n)$, one then finds that $\tau_{g,in}$ saturates at the value
\ba
\tau_{g,in}\approx \frac{\cal F}{3}\frac{b_{PR}}
{1-({\cal F}/3)\eta|\eta|}.
\label{eq:saturated_tau_g}
\ea 
This result and the whole picture of the orderly disk evolution are applicable only if 
\ba
{\cal F}<\frac{3}{\eta|\eta|}.
\label{eq:orderly_condition}
\ea

In the opposite case of runaway behavior, the first term in brackets in equation (\ref{eq:late}) rapidly comes to dominate over the second one.  In this case the equation then has an exponentially growing solution $\tau_{g,in}(\tilde t)\propto e^{\mu\tilde t}$, with a constant $\mu>0$ which satisfies the transcendental equation
\ba
1={\cal F}\eta|\eta|\frac{2-n}{4}\int\limits_0^{\infty}
e^{-\mu z}\tilde G\left(\frac{3(2-n)^2}{4}z\right)dz.
\label{eq:eq:trans}
\ea
It is easy to check that this equation has a positive solution (and the system evolves in a runaway fashion) only if 
\ba
{\cal F}>\frac{3}{\eta|\eta|},
\label{eq:runaway_condition}
\ea
which is the condition opposite to that in equation (\ref{eq:orderly_condition}).  In particular, if we assume $n=3/2$, which according to equation (\ref{eq:alpha}) corresponds to $T_{g}(r)=const$ profile in the gas disk, then equations (\ref{eq:eta_dim-less}) and (\ref{eq:Sig2}) imply that $\eta=7/4$, such that ${\cal F}>48/49\approx 1$ is needed for the disk to evolve in a runaway fashion. 

The upshot of this calculation is that, even in this simple model of the coupled disk evolution, runaway behavior becomes possible only when the feedback parameter ${\cal F}$ exceeds some critical value close to unity, given by equation
(\ref{eq:runaway_condition}).  We emphasize that despite the simplifications made in this section, the model presented is still fully global, as opposed to the treatment of \citet{Rafikov11b}.  Nevertheless, our result in equation (\ref{eq:runaway_condition}) is in agreement with the runaway condition ${\cal F}\gtrsim 1$ derived entirely from local considerations by \citet{Rafikov11b}.  And as we demonstrate in \S\ref{sect:results}, the general result given by equation (\ref{eq:runaway_condition}) remains valid even for the more complicated global models of the system, e.g.~in which the back-reaction of the aerodynamic drag on the gas disk is self-consistently included and the possibility that $\tau_g(R_{in})\lesssim 1$ is allowed.


\section{Solid-Gas Coupling in Eccentric Disks}
\label{app:ecc}

\begin{figure}
\resizebox{\hsize}{!}{\includegraphics[angle=0]{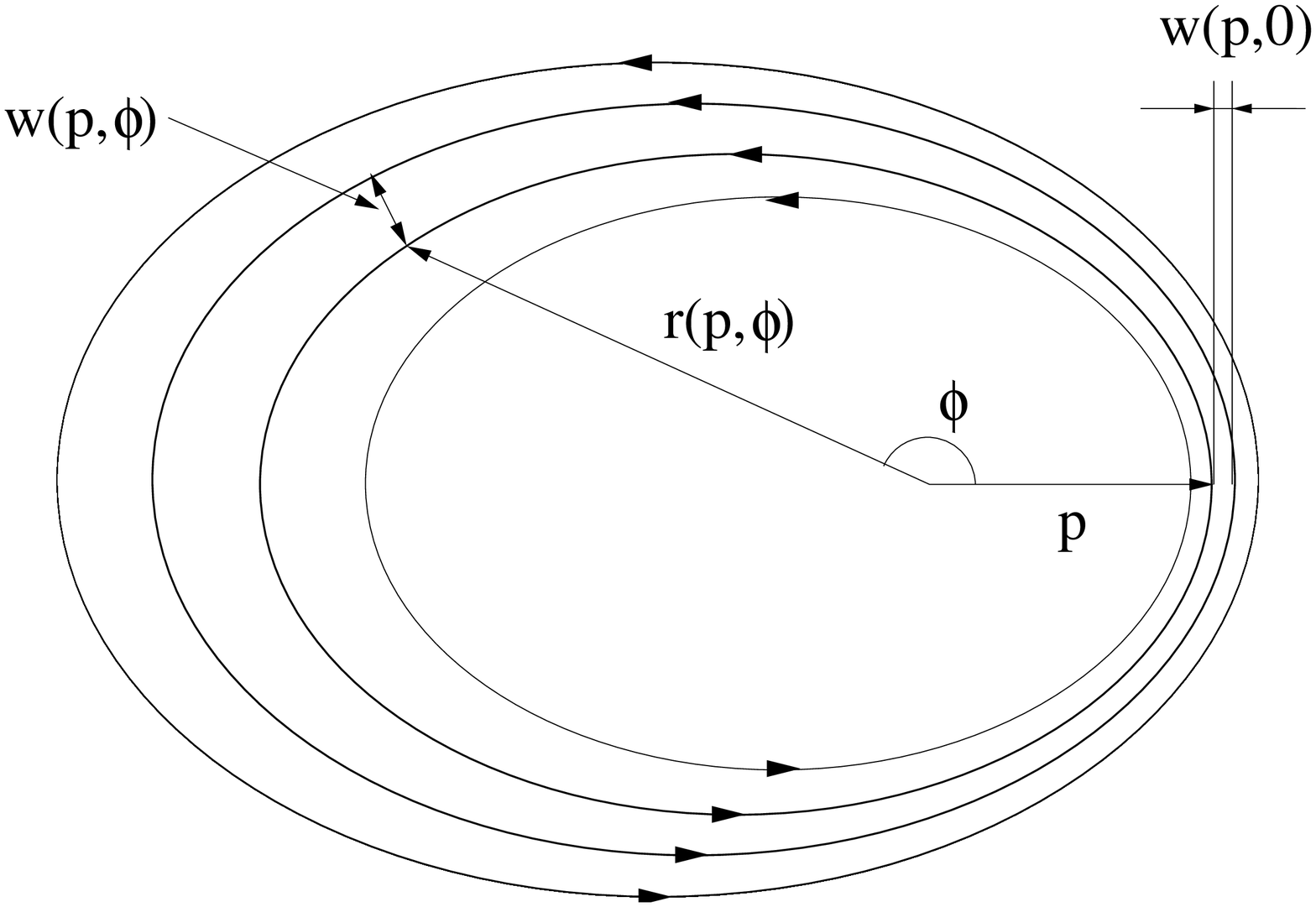}}
\caption[] {Schematic representation of the gas flow in an eccentric disk used to compute the aerodynamic drag between the solid and eccentric gaseous disks. See text for details.
}
\label{fig:eccentric_disk}
\end{figure}

We assume for simplicity that the streamlines in the gaseous disk are confocal ellipses (with foci coinciding with the WD position) with eccentricity $e$ being the same for all streamlines, and with apsidal lines pointing in the same direction (see Figure \ref{fig:eccentric_disk}).  There is no reason to believe that this geometric model captures all details of non-circular gas motion in real circum-WD disks --- we adopt it here just to facilitate our calculations.  Despite its simplicity, this model is sufficient to illustrate the importance of even moderately non-circular gas motions. 

We characterize each gas streamline by its periastron distance $p=a(1-e)$, where $a$ is the semi-major axis of a corresponding streamline.  Due to the non-circular shape of the streamlines, the surface density of gas varies as it travels along a given streamline. We characterize this variation by assuming that surface density is a function of $p$ and $\varphi$ --- the polar angle of the radius vector ${\bf r}$ at a given point on a streamline with respect to the apsidal line (see Fig.~\ref{fig:eccentric_disk}). The dependence of $\Sigma_g$ on $p$ characterizes the radial profile of the surface density. 

We denote $w(p,\varphi)$ as the (small) distance between the two neighboring streamlines (width of the flux tube confined by these streamlines) separated by the (small) distance $w(p,0)$ along the apsidal line. Clearly,
\ba
\frac{w(p,\varphi)}{w(p,0)}=|\nabla p(r,\varphi)|^{-1}=
\frac{1+e}{\left(1+2e\cos\varphi+e^2\right)^{1/2}},
\label{eq:w_rat}
\ea
where we have used the equation for elliptical orbit in polar coordinates 
\ba
p=r\frac{1+e \cos\varphi}{1+e}.
\label{eq:p}
\ea

The gas surface density at periastron $\Sigma_g(p,0)$ is related to the density $\Sigma_g(p,\varphi)$ at a different location along the same streamline via the continuity equation $\Sigma_g(p,\varphi)v(p,\varphi)w(p,\varphi)=
\Sigma_g(p,0)v(p,0)w(p,0)$, where 
\ba
v(p,\varphi)=\frac{\Omega(p) p}{1-e}
\left(\frac{1+2e\cos\varphi+e^2}{1-e^2}\right)^{1/2}
\label{eq:v_rat}
\ea
is the Keplerian velocity along the streamline (we neglect the modification of the gas rotation profile by pressure support).  Here $\Omega(p)$ is the value of $\Omega$ corresponding to the semimajor axis $a=p/(1-e)$.  Using equations (\ref{eq:w_rat}) and (\ref{eq:v_rat}) we obtain
\ba
\Sigma_g(p,\varphi)=\Sigma_g(p,0)\frac{(1+e)^2}
{1+2e\cos\varphi+e^2}.
\label{eq:Sigma_rat}
\ea

Now, let us consider a ring of particulate debris with radius $r$. An elliptic streamline passing through a point $(r,\varphi)$ on this ring has a periastron distance $p(r,\varphi)$ given by equation (\ref{eq:p}).  The azimuthal velocity of gas passing through $(r,\varphi)$ is given by 
\ba
v_{\varphi,g}(r,\varphi) &=& \left(\frac{GM_\star(1-e)}
{p(r,\varphi)}\right)^{1/2}\frac{1+e\cos\varphi}{\sqrt{1-e^2}}
\nonumber\\
&=&
\left(\frac{GM_\star}{r}\right)^{1/2}\left(1+e\cos\varphi\right)^{1/2},
\label{eq:vphig}
\ea
where we have again neglected pressure support in the gas disk.  The relative azimuthal velocity between the gas and particles at point $(r,\varphi)$ is then given by  
\ba
v_{\varphi,rel}=\left(\frac{GM_\star}{r}\right)^{1/2}
\left[1-\left(1+e\cos\varphi\right)^{1/2}\right],
\label{eq:v_rel2}
\ea
i.e. the gas rotates faster (slower) than particles at $\varphi=0$ ($\varphi=\pi$). 

The midplane gas density along the ring is given by $\rho(p,\varphi)= \Sigma_g(p,\varphi)\Omega(r)/c_s(r)$, where we have assumed for simplicity that $c_s$ is independent of $\varphi$.  According to equation (\ref{eq:aerodrag}) the azimuthal drag force acting per unit area of dense particle disk is then given by
\ba
f_{a}(r,\varphi)
&=& -\mbox{Re}_\star^{-1}\frac{GM_\star\Omega(r)}{rc_s(r)}\nonumber\\
&\times &\Sigma_g(p(r,\varphi),\varphi)
\left[1-\left(1+e\cos\varphi\right)^{1/2}\right]^2.
\label{eq:non_circ}
\ea

We now compute the orbit-averaged azimuthal force acting on the particle ring by averaging $f_{a}(r,\varphi)$ over $\varphi$.  To do this we will assume that $\Sigma_g(p,0)=\Sigma_{g,0}(p/p_0)^{-\theta}$, relate $\Sigma_g$ to $\Sigma_g(p,0)$ via equation (\ref{eq:Sigma_rat}) and use expression (\ref{eq:p}) 
to describe the dependence $p(r,\varphi)$.  As a result one finds that
\ba
\langle f_{a}(r)\rangle_\varphi &=&
-\mbox{Re}_\star^{-1}\frac{GM_\star\Omega(r)\Sigma_{g,0}}{rc_s(r)}
\left(\frac{p_0}{r}\right)^\theta I(e,\theta),
\label{eq:force_av}\\
I(e,\theta) &=& \frac{(1+e)^{2+\theta}}{2\pi}
\nonumber\\
&\times &
\int\limits_0^{2\pi}
\frac{\left[1-\left(1+e\cos\varphi\right)^{1/2}\right]^2}
{(1+e\cos\varphi)^{\theta}\left(1+2e\cos\varphi+e^2\right)}.
\label{eq:I_e_theta}
\ea 

We now explore the small-$e$ limit of our results. When $e\to 0$ one can easily show that $I(e,\theta)\to e^2/8$ independent of $\theta$, such that the orbit-averaged azimuthal force can be written as 
\ba
\langle f_{a}(r)\rangle_\varphi &=&
-\frac{e^2}{8\mbox{Re}_\star}
\frac{GM_\star\Omega(r)}{rc_s(r)}\Sigma_{g}(r).
\label{eq:force_av_small_e}
\ea 
We use this result in \S\ref{subsect:ecc} to demonstrate the enhancement of aerodynamic drag as the result of non-circular motions in the gaseous disk.

\end{appendix}

\end{document}